\documentclass{ptephy_v1}

\preprintnumber{XXXX-XXXX} 

\usepackage{lineno}
\usepackage{url}

\newcommand{\xmax}{$X_{\rm max}$}
\newcommand{\gcm}{g/cm$^2$}

\begin{document}


\title{Past, Present and Future of UHECR Observations}


\author{B.R. Dawson}
\affil{Physics Department, University of Adelaide, Adelaide, Australia}

\author{M. Fukushima} \affil{Institute for Cosmic Ray Research,
  University of Tokyo, Kashiwa, Chiba, Japan}

\author{P. Sokolsky}
\affil{Department of Physics \& Astronomy, University of Utah, Salt Lake City, Utah, USA \email{ps@cosmic.utah.edu}}


\begin{abstract}%
Great advances have been made in the study of ultra-high energy cosmic
rays (UHECR) in the past two decades. These include the discovery of
the spectral cut-off near $5\times10^{19}$\,eV and complex
structure at lower energies, as well as increasingly precise
information about the composition of cosmic rays as a function of
energy. Important improvements in techniques, including extensive
surface detector arrays and high resolution air fluorescence detectors,
have been instrumental in facilitating this progress. We discuss the
status of the field, including the open questions about the nature of
spectral structure, systematic issues related to our understanding of
composition, and emerging evidence for anisotropy at the highest
energies. We review prospects for upgraded and future observatories
including Telescope Array, Pierre Auger and JEM-EUSO and other
space-based proposals, and discuss promising new technologies based on
radio emission from extensive air showers produced by UHECR.
\end{abstract}

\subjectindex{ultra-high energy cosmic rays, extensive air showers, high-energy astrophysics}

\maketitle

\section{Introduction}
Cosmic rays were discovered a little over 100 years
ago~\cite{DiscoveryCosmicRays,KampertWatson2012}. They were called
``cosmic'' to distinguish them from the then equally mysterious ``X
rays'' emanating from laboratory instruments and particular
minerals. The increase in intensity of this cosmic radiation with
altitude made it clear that the sources were extraterrestrial.  Over
the next few decades, the grand questions were developed about their
origin, the extent of the energy spectrum, and their composition. It
took close to one hundred years to find the end of the remarkable,
approximately power law energy spectrum, at energies near 50 Joules
per nucleus.

We now know a great deal more, and at lower energies in significant detail
(as in the isotopic composition and gamma/electron/positron fluxes). 
But at the ``frontier'' energies of $>10^{18}$\,eV, where the flux is most
likely extra-galactic, the experimental tools, while greatly improved,
are still too imprecise. While our ability to measure energy with reasonable
certitude has improved dramatically, we attempt to measure the cosmic ray
composition with what is effectively a blunt instrument. The measurement of
arrival directions has also improved markedly, but there, Nature has been
unkind and only hints of cosmic ray anisotropy and sources have appeared.

One hundred years ago, the argument was whether the radiation came
from the Earth or Space, whether it was composed of charged or neutral
particles, and what could be inferred about its energy by measuring
``penetrating power''.   Now the arguments relate to the nature of
observed structures in the energy spectrum - the knees, ankles and
final cut-off, whether the composition is protonic, a mixture of p, He
and CNO group nuclei, or significantly heavy up to Fe, and how this
interplays with the spectral structure. Hints of departure from arrival direction isotropy come and
go and we fervently hold on to the most recent observation hoping that
this time the significance will strengthen with additional data and a
source, or sources, will finally be found. But except for the spectrum
and its structures, much of what we argue about is ephemeral and can
easily change with modification of hadronic models or a decrease in
statistical significance of a source. It is a hard fact that we still
do not know with any real certainty the origin of cosmic rays above a
few tens of GeV in energy.  What is it then that we do know, how well,
and what are the implications?  This is what will be discussed at
length in this volume. Our knowledge of our deficiencies also leads to
new ideas for better detectors and new programmatic approaches to fill
in needed extrapolations from accelerator data and ancillary
measurements. This too will be explored in subsequent pages.

In the present paper we address broadly the experimental status above
$10^{18}$\,eV and briefly describe the current status and the evolution
of ultra-high energy (UHE) detection techniques. The pioneering
Volcano Ranch, Haverah Park, SUGAR, Yakutsk and Akeno
arrays~\cite{NaganoWatson2000,Linsley1963,Allan1962,McCusker1963,Bell1974,Afanasiev1993,Nagano1984}
led to the major leaps forward represented by the AGASA, Fly's Eye,
HiRes, Auger and TA
experiments~\cite{Nagano1984,Bergeson1977,Baltrusaitis1985,Thomson2004,Abraham2004,Sokolsky2007}.
The early detectors led us to more precise formulations of the
questions we now ask.  The decades-long development of these second
and third generation detectors culminated in the reliable results with
well-understood energy and geometrical resolution that have given us
new hope that many of the puzzles presented by UHE cosmic rays may
soon be answered.  The current generation of detectors has brought
definitive confirmation on hints of an ankle structure above
$10^{17}$\,eV~\cite{Abbasi2005} and settled once and for all the reality
of a cut-off at energies between 4 and 6 $\times$
$10^{19}$\,eV~\cite{Abbasi2008,Abraham2008}.  The ``holy grail'' of UHE
cosmic ray physics has been found. It is striking that the suppression
is in the same range of energy as the prediction of the
Greisen-Zatsepin-Kuzmin (GZK) effect~\cite{greisen,zatsepin_kuzmin}
(depending somewhat on the assumed distribution of sources,
composition and the injection spectrum).

Improvements in the determination of extensive air shower (EAS)
profiles and the depth of shower maximum, \xmax{}, brought about by
either using stereo air fluorescence measurements (HiRes and
TA)~\cite{Bird1993,Abbasi2010,Abbasi2009} or by hybrid surface
detector and air fluorescence measurements (Auger and
TA)~\cite{AbuZayyad2000b,AbuZayyad2001,TA_Composition_Hybrid,Abraham2010,Auger_longXmax}
have reduced the reconstruction uncertainties in \xmax{} to near 10\,\gcm{}
with systematic uncertainties approaching this number. We have reached the
point where the experimental measurements are becoming more precise
than the theoretical underpinnings of the shower simulations used to
extract composition information. While data from the LHC in the
forward region at an equivalent energy of $10^{17}$\,eV are very
helpful in tuning the various hadronic
models~\cite{Engel2011,dEnterria2011}, there are significant issues in
extrapolating to p-nucleus and nucleus - nucleus interactions at much
higher energies ($10^{18}$ -
$10^{20}$\,eV)~\cite{dEnterria2011}. Currently the combined systematic
uncertainties for data and simulations make it difficult to reliably find the
mix of protons, He, CNO and Fe that would match the observed \xmax{}
distributions. What can be said, and this is a great accomplishment,
is that there is very little iron nucleus 
component~\cite{AugerMassMixtures,TA_Composition_Hybrid}. Why this
should be the case, given the relative stability of the iron nucleus as
it travels through the relic photon fields, and its relatively high
acceleration efficiency, is a puzzle which we are just now beginning
to confront.

\section{Achievements in the Era of Very Large Observatories}

\subsection{The previous generation of detectors}

At the beginning of the 21st century, three experiments were studying
the highest energy cosmic rays: the Yakutsk array, the Akeno Giant Air
Shower Array (AGASA), and the High Resolution Fly's Eye (HiRes).

The Yakutsk array in Russia had operated in various forms since 1967,
and had reached a maximum collecting area of 17\,km$^2$ 
around 1990.  Subsequently, it was reconfigured to study lower energy
cosmic rays, and today it has an area of 8\,km$^2$.  While
its focus has changed, analyses are still done on the data from high
energy showers already collected e.g. \cite{Ivanov2015}.

AGASA, located 100\,km west of Tokyo at an average altitude of 667\,m,
operated from 1990 to 2004 as a 100\,km$^2$ array consisting of over
one hundred scintillator detectors inter-connected by optical fibers
for timing measurements and data collection~\cite{agasa_nim}.  It
pioneered many of the techniques employed today in more modern
observatories, and produced important results on the UHECR energy
spectrum, anisotropy and mass composition~\cite{Shinozaki2004}.  Of
particular historical interest was the observation by AGASA of the
continuation of the cosmic ray spectrum beyond $10^{20}$\,eV, with no
sign of a flux suppression~\cite{Takeda2003}.

HiRes was the successor to the first successful air fluorescence
detector, the Fly's Eye~\cite{Baltrusaitis1985,YoshidaDai1998} which
operated from 1981 to 1993 at the Dugway Proving Grounds in Utah, USA.
The Fly's Eye achieved a time-averaged aperture of about 100\,km$^2$sr
at the highest energies, taking into account that it only operated on
clear, moonless nights.  HiRes was an advancement in resolution and
sensitivity, achieved by increasing the telescope effective mirror
areas to 3.8\,m$^2$, and reducing the camera pixel angular diameters to
$1^\circ$~\cite{AbuZayyad2000a}.  Two sites, 12.8\,km apart, were
instrumented, allowing for stereo observations of approximately
$30-40$\% of air showers that triggered either detector near
$10^{19}$\,eV.  The collecting area of HiRes was close to an order of
magnitude larger than that of the Fly's Eye.  The first HiRes site at
Five Mile Hill began full operation in 1997, followed by the Camel's
Back Mountain site in 1999.  HiRes ceased operations in 2006.  A
summary of the important physics results from HiRes is given
in~\cite{Sokolsky2007}.  This includes the first unambiguous detection
of a flux suppression at the highest energies, using monocular data
from HiRes I and published in 2008~\cite{Abbasi2008}.

At the beginning of the 1990s discussions began about the next step in
UHE cosmic ray observations, where it was recognised that apertures
even larger than that of HiRes would be necessary to answer some of
the long-standing questions in cosmic ray astrophysics.  The
experimental challenge was enhanced by the apparent disagreement in
the energy spectra presented by AGASA and HiRes in the first few years
of the new century. This led to the next generation of experiments
adopting hybrid designs, with combinations of surface arrays and air
fluorescence detectors.

\begin{figure}[!t]
\centering
\begin{minipage}{.48\textwidth}
  \centering
  \includegraphics[width=0.95\linewidth]{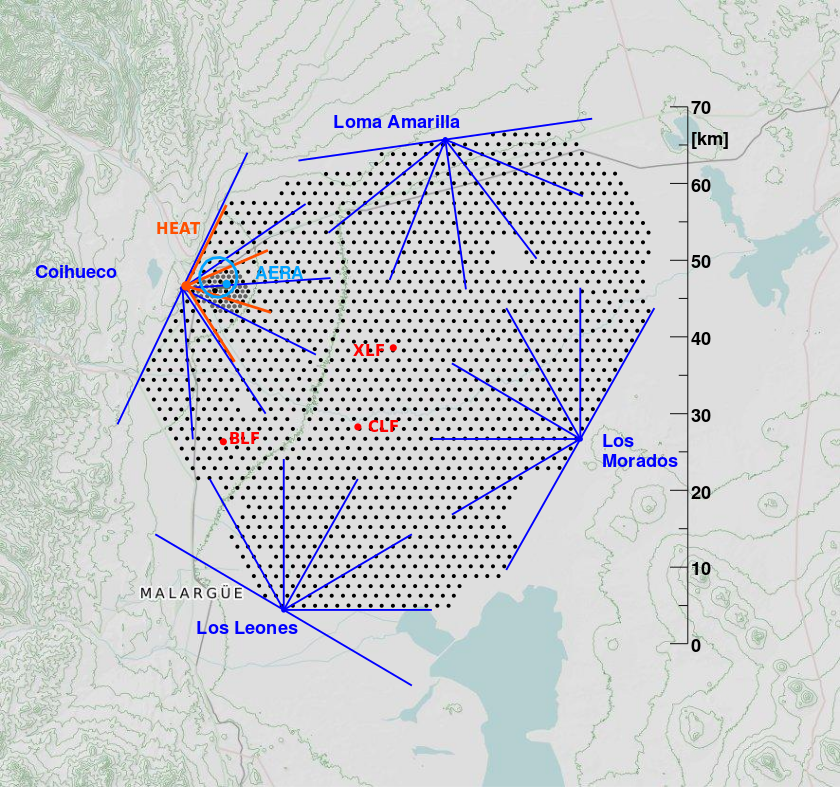}
  \caption{Layout of the Pierre Auger Observatory, showing the
    positions of the 1660 SD stations, the fields of view of the main FD telescopes (in blue) and the fields of view of the HEAT high-elevation telescopes (in red).  For further details see~\cite{AugerNIM}.}
  \label{fig:Auger}
\end{minipage}%
\hfill
\begin{minipage}{.48\textwidth}
  \centering
  \includegraphics[width=0.95\linewidth]{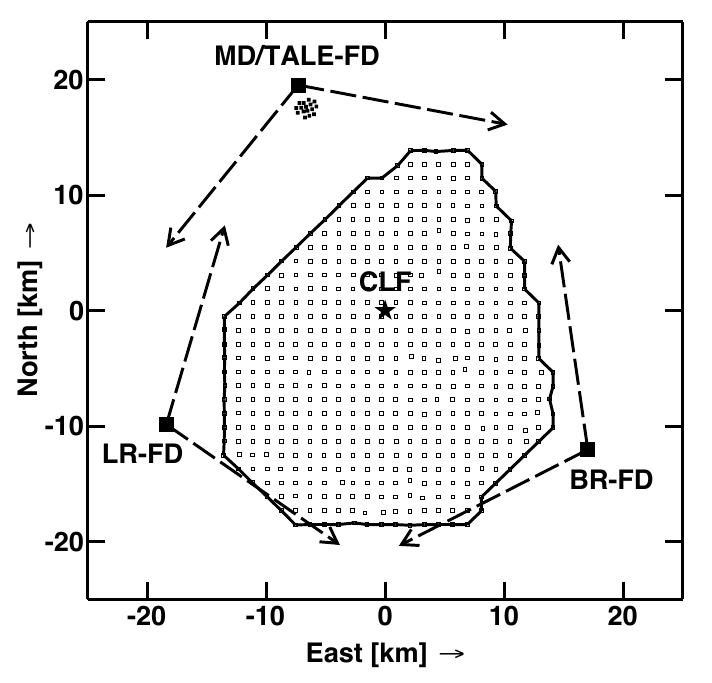}
  \caption{The Telescope Array layout, showing the locations of the
    507 surface detectors.  The three FD stations at Middle Drum,
    Black Rock Mesa and Long Ridge are indicated, each having an azimuthal
    field of view of about $110^\circ$.  The TALE detector is situated
    at the MD site~\cite{TA_Spectrum_ICRC15}.}
  \label{fig:TA}
\end{minipage}
\end{figure}

\subsection{Currently operating Observatories}

\subsubsection{The Pierre Auger Observatory}
\label{Sect2:auger}

The Pierre Auger Observatory had its beginnings in 1991 when James
Cronin and Alan Watson began discussions with a number of
experimenters in the field.  Its design evolved through an initial
meeting in Paris in 1992, a two week design workshop in Adelaide in
1993, and a six-month design study hosted by Fermilab in 1995.  Initial
ideas were based on a 5000\,km$^2$ surface array without fluorescence
telescopes, but the advantages of the hybrid approach soon became
apparent~\cite{Watson2014,SommersHybrid,DawsonHybrid}.  The Auger Observatory
is near Malarg\"{u}e, Argentina, and its construction began in 2001
with an engineering array.  The Observatory was completed in 2008,
though official data taking began in 2004 during construction.

Auger's surface detector (SD) consists of 1660 water-Cherenkov
detectors (WCDs) arranged on a 1.5\,km triangular grid covering
3000\,km$^2$~\cite{AugerNIM}, see Figure~\ref{fig:Auger}.  The WCDs
are 10\,m$^2$ in area and 1.2\,m deep, and build on the experience
gained from the Haverah Park detector in the UK (see
~\cite{KampertWatson2012}). Such detectors have the advantage of
having a broad zenith angle sensitivity, and are deep enough to
produce signal from the numerous photons in the extensive air shower.
The SD is fully efficient for cosmic ray energies greater than $
3\times 10^{18}$\,eV and zenith angles less than $60^\circ$.  A small
23.5\,km$^2$ area of the array hosts a denser 750\,m spacing of WCDs,
which is fully efficient for $E > 3 \times 10^{17}$\,eV and zenith
angles less than $55^\circ$.

The main fluorescence detector (FD) consists of 4 sites on the
perimeter of the surface array.  Each site hosts six telescopes, each with a
field of view of $30^\circ$ in azimuth and an elevation range from
$1.5^\circ$ to $30^\circ$.  Each telescope is of the Schmidt design and
consists of a 13\,m$^2$ segmented spherical mirror with a 2.2\,m
diameter entrance aperture (including a ring of corrector lenses) and
a camera composed of 440 photomultiplier pixels, each viewing a
$1.5^\circ$ diameter area of sky~\cite{AugerNIM}.  The entrance
aperture also contains a UV transmitting filter to match the air
fluorescence spectrum spanning approximately 300--400\,nm.  The
telescope design benefits from experience with the Fly's Eye and HiRes
experiments, with the primary difference being the Schmidt optics
design, allowing for a wide field of view with minimal coma
aberration.  Apart from the 24 telescopes of the main FD described
here, an additional three telescopes make up the HEAT (High Elevation
Auger Telescopes) system.  These view the elevation range between
$30^\circ$ and $58^\circ$ to study lower energy air showers (currently
down to $10^{17}$\,eV) that, due to their lower brightness, are
observed closer to the FD site~\cite{AugerNIM}.

The combination of the surface and fluorescence detectors to make a
``hybrid'' observatory has been exploited in much of Auger's
scientific output.  The SD has many strengths, including robust WCDs
that operate with a 100\% duty cycle.  It also measures the lateral
characteristics of the air shower, albeit at one altitude, which are
being used for several studies including mass composition.  The FD,
while having the disadvantage of a 15\% duty cycle~\cite{AugerNIM},
measures fluorescence light which is produced in direct proportion to
the energy deposited by the air shower.  Thus, the fluorescence
technique measures air shower energies calorimetrically, and it is the
FD measurements of energy that calibrate the SD energy scale, as
described below.  The FD views the developing shower and has access to
the depth of shower maximum, $X_{\rm max}$, used in mass composition
studies.  Finally, the FD reconstruction of the direction and position
of the air shower axis is greatly assisted by the SD measurement of
the shower arrival time at ground level~\cite{DawsonHybrid}.  This
hybrid reconstruction produces an FD arrival direction resolution of
about $0.5^\circ$~\cite{Bonifazi2009}, which helps achieve typical
resolutions in energy and $X_{\rm max}$ of $\sim10$\% and
20\,g/cm$^2$, respectively, at $10^{19}$\,eV~\cite{Dawson2007}.

Descriptions of the calibration procedures for both the SD and FD are
given in ~\cite{AugerNIM}.  In both cases the calibration is
``end-to-end'', either using unaccompanied muons (in the case of the
SD) or a large ``drum'' to illuminate an FD aperture, so as to
calibrate the full detector and data acquisition chain in one step.
The atmosphere is also carefully monitored.  The density of the lower
atmosphere has a well known effect on the lateral distribution of the
air shower at ground level, and these weather corrections are applied
to the SD energy measurements for certain studies, such as large scale
anisotropy measurements~\cite{Auger_weather2016}.  Finally, the light
attenuation characteristics of the atmosphere are measured on an
hourly basis during FD operations to account for varying molecular and
aerosol scattering, and to monitor cloud cover~\cite{AugerAtmosphere2010}.

The Auger Collaboration had always planned to build a northern array
in order to achieve full sky coverage. The site was chosen to be in
south-eastern Colorado, USA~\cite{AugerNorth}.  Currently there is a
strong focus on an upgrade of the southern site~\cite{AugerPrime}
(discussed in Section 4), and exploration of the northern sky 
is being undertaken by the Telescope Array collaboration.

\subsubsection{The Telescope Array}

The Telescope Array (TA) project was originally proposed around 1997
by members of the AGASA and HiRes experiments as a large fluorescence
telescope complex with an effective aperture (after accounting for a
10\% duty factor) of 5,000\,km$^2$ sr for 10$^{20}$~eV cosmic ray
particles~\cite{ta_durban_1, ta_durban_2, ta_dr_2000}.
An apparent discrepancy in the measurements of the UHECR flux above
the GZK energy by AGASA and HiRes (see \cite{Takeda2003,Abbasi2008}),
however, encouraged the members of TA to make a critical examination
of the experimental methods used. This led to the present form of TA,
started in 2003, as two complete and co-sited SD and FD detectors,
each observing the same UHECR events and allowing for a critical
comparison of the measured shower parameters~\cite{ta_tsukuba}.  The TA
experiment occupies a large area near the town of Delta, 200\,km
south-west of Salt Lake City, Utah, USA, and is now operated by a
collaboration of 120 members from five countries: Japan, USA, Korea,
Russia and Belgium. The experiment is conducting a high-statistics
exploration of the northern sky as a hybrid array of surface and
fluorescence detectors.

The TA surface detector comprises 507 detectors on a 1.2\,km
square grid covering an area of 700\,km$^2$ (Figure~\ref{fig:TA}).
Each detector has an area of 3\,m$^2$ and consists of two layers of
1.2\,cm plastic scintillator separated by a 1\,mm thickness of
stainless steel~\cite{TA_SD}.  The scintillators are equally sensitive
to all minimum ionising charged particles ($e^\pm$, $\mu^\pm$), with
the SD energy determination being dominated by the EM
(electromagnetic) component.  This is seen as an advantage as it
reduces uncertainties in the energy scale due to mass composition or
hadronic physics.  The array reaches full efficiency above $10^{19}$\,eV
for zenith angles less than $45^\circ$, providing an aperture of
1100\,km$^2$\,sr.

Three fluorescence detector sites sit near the boundary of the
array.  One of the sites, at Middle Drum, uses 14 refurbished telescopes
originally part of the HiRes detector.  They are arranged in two
``rings'' that together view an elevation range from $3^\circ$ to $31^\circ$,
and an azimuth range of $112^\circ$.  Each telescope consists of a
2\,m diameter spherical mirror and a camera of 256 hexagonal pixels, with
pixels viewing approximately a $1^\circ$ diameter section of the
sky~\cite{AbuZayyad2000a}.

The other two FD sites at Black Rock Mesa (BR) and Long Ridge (LR)
each contains 12 newly fabricated telescopes.  Each telescope consists
of a 3.3\,m diameter segmented spherical mirror focusing light onto a
256 pixel camera.  The pixels also have a field of view of $1^\circ$
diameter and each site covers a field of view of $3^\circ - 33^\circ$
in elevation and $108^\circ$ in azimuth using a two ring
structure~\cite{TA_newFD}.  The electronics in the new FD stations
digitise the pixel signals at 10\,MHz with 14 bits of
precision~\cite{TA_FD_electronics}.

The TA detectors began full operation in March 2008.  Various data
sets are being collected, and the consistency between different data
sets and different analysis procedures has been carefully examined.  Surface
detector energy spectrum studies use contained events within the array
with zenith angles $< 45^\circ$, while anisotropy studies use looser
cuts and zenith angles $< 55^\circ$~\cite{Tinyakov2014}.  For
fluorescence analysis, some studies are done with mono observations
(requiring observations from one FD site), stereo observations (two FD
sites), and hybrid observations.  In the latter case, FD observations
are coupled with SD measurements of the shower at ground level, much
in the style of Auger analysis, except that timing information from
more than one SD station is used, and the shower core location derived
from SD data alone is used to constrain the hybrid
core~\cite{TA_Composition_Hybrid}.

The atmospheric transparency of the TA site is monitored by a suite of
instruments, including a Central Laser Facility, IR cloud cameras and
a LIDAR station~\cite{tomida_2011,Tomida2013}.  In analyses performed
to this point, the aerosol content of the atmosphere has been assumed
to be constant with time.  Atmospheric transparency data has been used
to determine an average value of the vertical aerosol optical depth of
0.04 (see e.g. ~\cite{TA_Composition_Hybrid}), and the effect on
systematic uncertainties of fluctuations about the mean has been studied.

The TA project has in recent years extended its reach towards lower
energy cosmic rays with the TALE (TA Low Energy) extension.  With an
additional 10 telescopes in an extra two ``rings'' at the Middle Drum
site, the field of view there is now $3 - 59^\circ$ in elevation and
approximately $120^\circ$ in azimuth.  By using Cherenkov-rich events,
the energy threshold of TALE is below $10^{16}$\,eV, complementary to
the standard fluorescence observations~\cite{Zundel_ICRC15}.  The
surface detector array is being increased in density in front of this
FD site to assist with lower energy hybrid observations.

\subsection{Advances in techniques}
\label{Sect2:advances}

The advances in our understanding of UHECR that we will discuss in the
following sections owe much to the large collecting areas now
instrumented by the Auger and TA collaborations.  Additionally, the
number of scientists now studying UHECR is much larger than in early
generations of experiments, meaning that more manpower is available
for the maintenance of the experiments, for calibration of the
detectors and the atmosphere, and for devising new and creative
analysis techniques and cross-checks.  Thus the increase in
sensitivity of the modern observatories is due to more than an
increase in the collecting area alone.  We give some examples here of
recently exploited advances in detectors, tools and techniques.

A stable surface detector is necessary for optimal energy resolution,
and for searching for weak broad-scale anisotropies.  For example,
temperature-dependent particle density measurements can introduce
diurnal variations into shower rates above some energy threshold,
which may be wrongly interpreted as sidereal harmonics.  Both Auger
and TA avoid this by monitoring their SD detector performance on short
time scales.  This is done by collecting histograms of the fundamental
unit used to calibrate SD signals - in the case of TA scintillator
detectors, a histogram of integrated charges from $\sim$0.4 million
penetrating particles is collected every 10 minutes for each
detector~\cite{TA_SD}, while for Auger WCDs, histograms of signals
from through-going muons are collected every minute, from which the
signal due to a ``vertical-equivalent muon'' (VEM) can be
derived~\cite{AugerNIM}.  In addition to this basic calibration, Auger
also makes a correction (in broad-scale anisotropy studies) for the
effect of diurnal atmospheric variations on air-shower development.
These weather effects make small but significant corrections to the SD
energy estimator $S(1000)$ (the detector signal 1000\,m from the
shower core) using the local air density and
pressure~\cite{Auger_weather2016}.

In fluorescence light detection, both collaborations have benefitted
from experience with the HiRes and Fly's Eye experiments.  The
electronics in the new FDs in both TA~\cite{TA_newFD} and
Auger~\cite{AugerNIM} include sophisticated triggering circuitry, and
digitisation of each pixel signal is performed at 10\,MHz.  Improved
telescope design is a feature in both experiments, with Auger using a
Schmidt optics design which gives a coma-free optical spot over a
$30^\circ$ field of view~\cite{AugerNIM}.  At TA, signals from
UV-bright stars are used to verify ray-tracing estimations of optical
aberrations, and to check telescope alignment~\cite{TA_newFD}.
Calibrations of the FDs feature end-to-end procedures.  At Auger, a
``drum'' calibrating system is moved from telescope to telescope
periodically.  It uniformly illuminates the aperture of a telescope
with an absolutely calibrated light source at a number of wavelengths.
Between these absolute drum calibrations, the system calibration is
monitored with light sources illuminating the mirror and
camera~\cite{AugerNIM}. At TA, a small number of absolutely-calibrated
PMTs are in each camera~\cite{kawana_crays}, and their gains are
monitored with a radioactive source-scintillator YAP
unit~\cite{shin_yap}.  The other PMTs in the camera are
cross-calibrated and monitored every 30 minutes during observation
time using a diffuse Xenon light source installed in front of each
camera~\cite{TA_FDcalib}.  The Telescope Array is going one step
further, experimenting at the Black Rock FD station with an electron
light source (ELS), which shoots a vertical beam of 40\,MeV electrons
100\,m from a FD telescope, creating artificial air fluorescence
in-situ~\cite{shibata_els}.  The ELS aims to make an end-to-end
calibration of the FD - from energy deposits in the air to the
detection of fluorescence light by the telescope~\cite{TA_ELS_ICRC15}.

Since both experiments rely on fluorescence measurements to calibrate
the SD energy scales, much effort has recently gone into laboratory
measurements of the fluorescence efficiency, that fraction of the
shower's ionisation energy deposit going into light production.  After
the pioneering work of Bunner, Kakimoto et al. and Nagano et
al.~\cite{Bunner,Kakimoto,Nagano}, new measurements include those of
the AIRFLY~\cite{AIRFLY} and FLASH~\cite{FLASH} experiments.  Results
include precise measurements of the fluorescence efficiency and
spectrum, and the pressure, temperature and humidity dependence of the
light~\cite{AIRFLY}.  Auger, which uses the AIRFLY results, has been
able to reduce the systematic uncertainty in shower energy associated
with the fluorescence yield from 14\% to
3.6\%~\cite{Auger_EScale_ICRC13}.  TA has used the Kakimoto et
al.~\cite{Kakimoto} fluorescence yield model, also used by HiRes, for
the sake of consistency and continuity.  It is now re-examining its
energy scale using more contemporary measurements.

Air shower simulations are used in a variety of applications in both
experiments, including the extraction of mass composition estimates
from air shower development measurements.  The CORSIKA three
dimensional shower simulation program~\cite{CORSIKA} is still the most
widely used, with other code such as the longitudinal profile
simulator CONEX~\cite{CONEX} being used in certain applications.  The
continuing challenge is to improve the implementation of high energy
hadronic interactions in the code.  In the last decade important new
constraints on these interactions have come from measurements at the
Large Hadron Collider, spawning new hadronic models for CORSIKA and
CONEX, including EPOS-LHC~\cite{EPOS}, QGSJetII-04~\cite{QGSJet} and
Sibyll~2.3~\cite{Sibyll}.  Information is also moving in the other
direction, with cosmic ray experiments like Auger and TA testing
certain aspects of these predictions (see
Section~\ref{Sect2:interactions}).

\subsection{The energy spectrum of UHECR}
\label{Sect2:energy}

The surface detectors of Auger and TA are the ``work-horses'' of the
respective experiments, since they are operational 24 hours per day
and contribute the majority of exposure to studies such as the energy
spectrum.  However, both experiments use their fluorescence detectors to
determine the energy scale of their surface detectors, avoiding much
of the uncertainty associated with the alternative, air shower
simulations.  Systematic uncertainties in the energy scale have been
derived taking account of uncertainties in detector calibration and
stability, atmospheric transmission, fluorescence yield, and
reconstruction.  For Auger this amounts to a total systematic
uncertainty of 14\%~\cite{Auger_EScale_ICRC13}, and for TA
21\%~\cite{TA_hybrid_spectrum2015}.  Both quoted uncertainties are
independent of energy.

\begin{figure}[!t]
\centering
\begin{minipage}{.48\textwidth}
  \centering
  \includegraphics[width=1.0\linewidth]{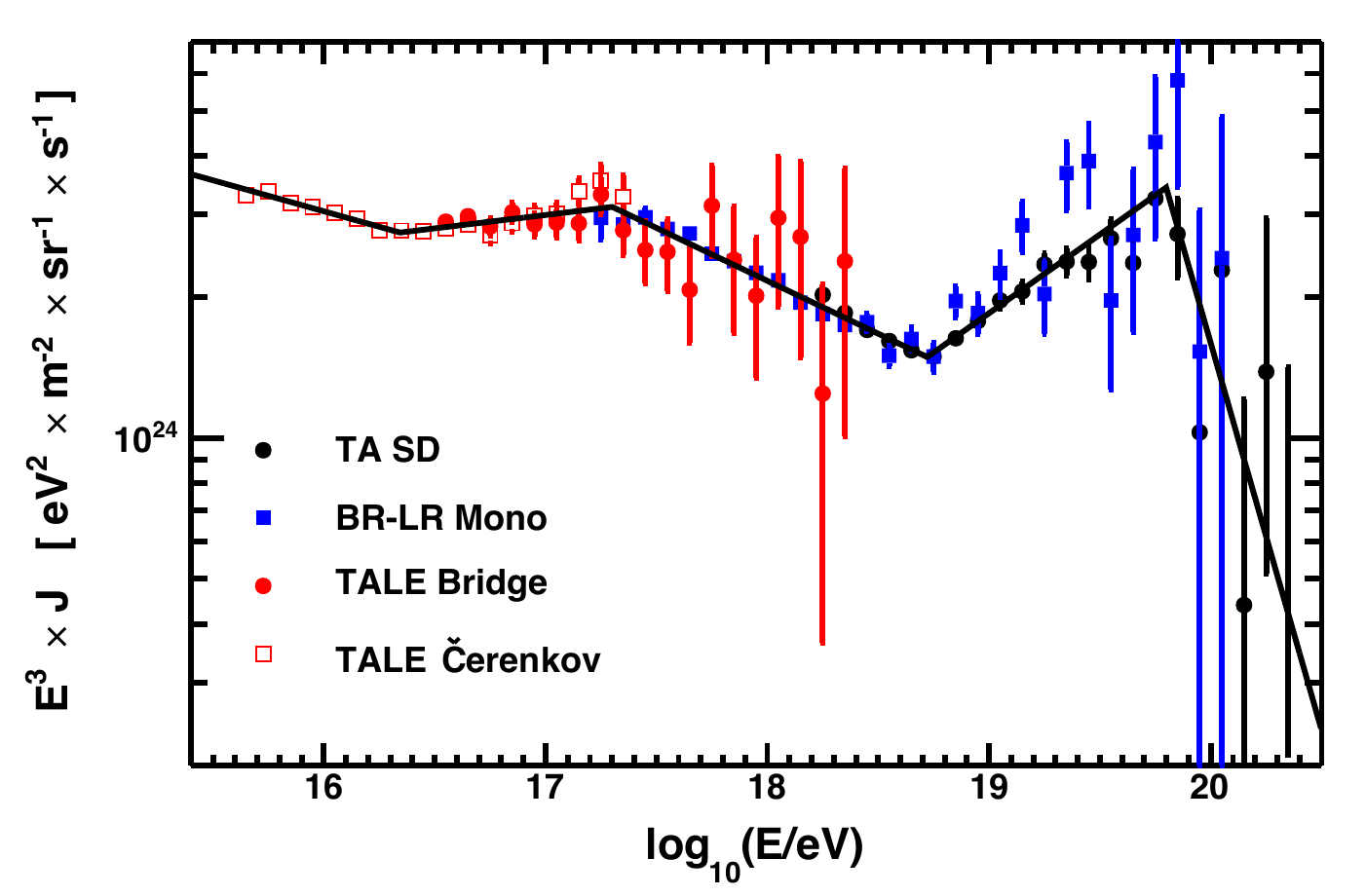}
  \caption{TA energy spectrum (flux scaled by $E^3$) with
    data from TALE, the LR and BR FDs, and the surface
    detector~\cite{TA_Spectrum_ICRC15}. Five power laws are fit to
    indicate the spectral structure.  The systematic uncertainty on the energy scale is 21\%.}
  \label{fig:TASpectrum}
\end{minipage}%
\hfill
\begin{minipage}{.48\textwidth}
  \centering
  \includegraphics[width=1.0\linewidth]{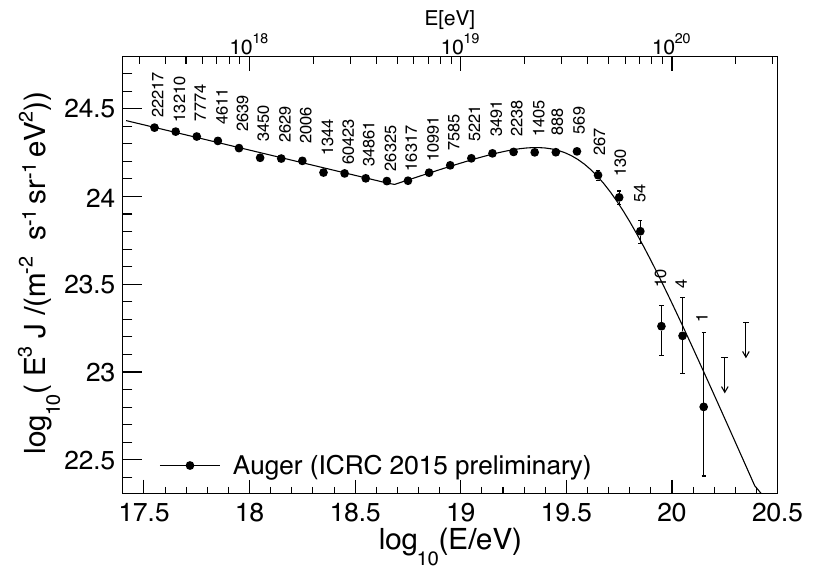}
  \caption{The Auger combined energy spectrum with data from the
    750\,m spaced SD, FD (hybrid) and the 1500\,m SD array.  The
    energy systematic uncertainty is 14\%.  Event numbers are
    shown, and a spectrum model is
    fitted~\cite{Auger_Spectrum_ICRC15}.}
  \label{fig:AugerSpectrum}
\end{minipage}
\end{figure}

The most recent energy spectra were presented in 2015 by
Auger~\cite{Auger_Spectrum_ICRC15} and TA~\cite{TA_Spectrum_ICRC15}.
The Telescope Array presented a spectrum over 4.5 decades of energy
from below $10^{16}$\,eV, combining results from the SD, the two
Japanese FD sites (monocular reconstruction) and TALE
(Figure~\ref{fig:TASpectrum}).  At $10^{20}$\,eV the exposure is
approximately 6200\,km$^2$\,sr\,yr for the SD and 650\,km$^2$\,sr\,yr
for the FDs.  A remarkable set of features is present in the combined
spectrum, represented by a series of five power-law segments.  The
flux suppression at the highest energies, identified by TA as
consistent with the GZK mechanism~\cite{greisen,zatsepin_kuzmin}, is
present above $10^{19.80\pm0.05}$\,eV ($6.3\times 10^{19}$\,eV), and
the spectral ankle appears at $10^{18.72\pm0.02}$\,eV ($5.2\times
10^{18}$\,eV), again consistent with proton interactions with the
cosmic microwave background (CMB)~\cite{SpectralDip}.  At lower energies,
TALE detects two other features, a second knee at
$10^{17.30\pm0.05}$\,eV ($2.0\times 10^{17}$\,eV), and a low energy
ankle at $10^{16.34\pm0.04}$\,eV ($2.2\times 10^{16}$\,eV).  The
systematic uncertainties of the TALE measurements are currently being
evaluated.

The latest Auger spectrum~\cite{Auger_Spectrum_ICRC15}
(Figure~\ref{fig:AugerSpectrum}) is a combined measurement from the
1500\,m spaced SD (zenith angle $\theta < 80^\circ$), the FDs
operating in hybrid mode, and the 750\,m spaced SD ($\theta <
55^\circ$).  Two separate analyses were performed for the 1500\,m SD
array, for ``vertical'' ($\theta < 60^\circ$) and for ``inclined''
($60^\circ < \theta < 80^\circ$) showers, with the latter events being
muon dominated.  The integrated exposures for the 1500\,m vertical SD,
inclined SD and 750\,m SD are 42,500\,km$^2$\,sr\,yr,
10,900\,km$^2$\,sr\,yr and 150\,km$^2$\,sr\,yr, respectively.  The
energy-dependent exposure for the hybrid spectrum is
1500\,km$^2$\,sr\,yr at $10^{19}$\,eV.  The combined spectrum extends
from $10^{17.5}$\,eV to the highest energies, and shows the ankle
feature at $(4.82 \pm 0.07 \pm 0.8) \times 10^{18}$\,eV (statistical
and systematic uncertainties are quoted).  The flux suppression is
characterised by a smooth function with $E_s = (4.21 \pm 0.17 \pm
0.76)\times 10^{19}$\,eV, with $E_s$ representing the energy at which
the flux falls to one-half of the value of the power-law
extrapolation.

When comparing the Auger and TA spectra, the following points have
been made by a joint Auger/TA energy spectrum working
group~\cite{UHECR2016_spectrum}:

\begin{itemize}
  \item In the overlapping region of energy, the spectral slopes are
    consistent within uncertainties, and the energy of the ``ankle'' is
    consistent given the statistical and systematic uncertainties.  A flux
    difference at energies from $10^{17.5} - 10^{19.3}$\,eV of $\sim
    20\%$ could be the result of a shift in the energy scale within
    systematic uncertainties.

  \item On the other hand, the energy of the flux suppression at the
    highest energies, characterised by $E_{1/2}$ (a measure of the
    suppression energy favoured by TA~\cite{Ehalf}) is inconsistent.  The TA
    measurement is $(6.0 \pm 0.7 {\rm (stat)}) \times 10^{19}$\,eV,
    compared with the Auger measurement of $(2.47 \pm 0.01 {\rm
      (stat)} ^{+ 0.82}_{- 0.34} {\rm (syst)}) \times 10^{19}$\,eV.

  \item The agreement in the position of the ankle and the
    disagreement in the suppression energy might be explained by an
    energy-dependent systematic uncertainty in energy, or a real
    difference in the physics of cosmic rays in the northern and southern
    hemispheres.  At the current time, no source of the former has
    been identified and, as mentioned above, both experiments quote an
    energy independent systematic.  As an example, differences in the
    correction for {\em invisible energy} used by Auger and TA in the FD
    analyses (i.e. that energy carried by high energy muons and
    neutrinos that does not result in proportionate fluorescence
    light), and differences in fluorescence yield models, produce only
    a small shift in the energy scale of $5-10$\% which is essentially
    energy independent.

    \item The possibility that the UHECR sky is different in the
      northern and southern hemispheres has been studied by
      determining the energy spectrum as a function of declination.
      The Auger SD ``vertical'' ($\theta < 60^\circ$) spectrum covers
      a declination range from $-90^\circ$ to $+25^\circ$, 71\% of the
      total sky.  Four energy spectra have been derived for
      independent declination bands~\cite{Auger_Spectrum_ICRC15},
      which are then compared with the total Auger spectrum.  A small,
      and statistically insignificant declination dependence in the
      flux is observed ($< 5$\% below the suppression energy $E_s$ and
      $<13$\% above) within the declinations studied.  The conclusion
      is that the Auger/TA spectrum difference in the suppression
      region cannot be explained in terms of a declination dependence,
      unless there is a significant change in the spectrum north of
      $25^\circ$.  The TA collaboration see a hint of such an effect
      when considering an SD energy spectrum extended to include
      zenith angles $< 55^\circ$, covering a range of declinations
      from $-16^\circ$ to $+90^\circ$.  The position of the
      suppression energy $E_{1/2}$ is approximately $3\sigma$ higher
      for declinations north of $26^\circ$ compared with those south
      of $26^\circ$~\cite{Verzi_PTEP}.  The question of a declination
      dependent spectrum is connected to the observations of
      anisotropies of the flux discussed below in
      Section~\ref{Sect2:anisotropy}.

\end{itemize}

Currently the source of the disagreement in the energy spectrum at the
highest energies is an open question.  Any declination-dependence of
the spectrum appears weak, but more studies are on-going.  In
parallel, the Auger and TA groups are working together to understand
differences in the analysis procedures, and how these might lead to an
experimental explanation for the spectral differences.

The methods used by the collaborations for measuring the energy
spectrum have many things in common, most importantly in the use of
fluorescence measurements to set the energy scale.  But there are
significant differences in other areas, either necessitated by
detector differences (scintillators vs. water-Cherenkov detectors), or
because of the philosophy of the collaborations.  A case in point is
how each experiment accounts for the zenith angle-dependent
attenuation of showers for SD measurements.  The Auger collaboration
has a philosophy of avoiding the use of air shower simulations,
wherever possible, in deriving energy estimates.  To account for
shower attenuation, Auger uses the method of constant intensity cuts
(CIC), a data-driven procedure which uses the fact that the intensity
of cosmic rays above a certain energy threshold should be independent
of the zenith angle of the showers~\cite{AugerNIM}.  The analysis
converts Auger's SD energy estimator $S(1000)$ (the WCD signal 1000\,m
from the shower core) to $S_{38}$, the value of $S(1000)$ the shower
would have possessed if it had arrived at the median zenith angle of
$38^\circ$.  While, in principle, the conversion from $S(1000)$ to
$S_{38}$ could be energy-dependent, no dependence has been detected.
In contrast, the TA analysis uses simulations of proton showers to
account for shower attenuation in their analysis.  Both methods are
valid, but both have possible weaknesses, which can be explored in
future studies under the joint working group structure created by the
two collaborations~\cite{UHECR2016_spectrum}.

The astrophysical interpretation of the energy spectrum is, of course,
coupled to other measurements made by the collaborations, in
particular the mass composition.  The TA collaboration finds that
features of its spectrum can be satisfactorily explained by models
of production and propagation of a pure protonic cosmic ray
flux~\cite{TA_ICRC15_spectrum_interpretation}.  Here, the ankle is
interpreted as a ``dip'' caused by pair production on the CMB and IR
photons, and the suppression is due to the classic
Greisen-Zatsepin-Kuzmin photopion production on the same photon
fields.  Propagation simulations for both a uniform distribution of
proton sources, and a distribution which follows the local large scale
structure of the Universe, are compared with the measured spectrum to
fit the power-law index $\gamma$ of the spectrum at the sources, a
parameter $m$ related to the source evolution with redshift, and a
logarithmic shift of the experimental energy scale $\Delta \log E$.
Good fits were obtained for both source distributions. For the uniform
distribution, the $\chi^2/d.o.f.$ was 12.4/17 with $\gamma =
2.21^{+0.10}_{-0.15}$, $m = 6.7^{+1.7}_{-1.4}$, and $\Delta
\log E =0.03\pm0.03$~\cite{TA_ICRC15_spectrum_interpretation}.  The
assumption of the pure proton flux is consistent with TA's
measurements of mass composition (see Section~\ref{Sect2:mass}).  The
best value of energy shift ($\sim 3$\%) is well within the systematic
energy uncertainty of TA. The parameters $\gamma$ and $m$ apply only
for sources with $z < 0.7$, since the contribution of protons arriving
from sources beyond that redshift is negligible for $E >
10^{18.2}$\,eV.

Such a mass composition is not favoured by Auger's measurements.  In
this case it is necessary to use source production and propagation
modelling to fit both the energy spectrum, and the mass composition
measured at Earth.  This has been done by several authors, including
the Auger collaboration~\cite{Auger_ICRC15_spectrum_interpretation}.
The input to the simulation is a population of uniformly distributed
sources accelerating protons, and nuclei of He, N and Fe, to a maximum
rigidity with a power-law spectrum of index $\gamma$.  The standard
interactions of protons and nuclei with background photons are taken
into account.  The result of this simple model is a rather hard input
spectrum ($\gamma \sim 1$) with a rather low maximum rigidity of the
source accelerators.  While the authors point out the naivety of the
model, the results are in real contrast to the protonic model favoured
by the TA collaboration, and thus stress the importance of the mass
composition assumptions when interpreting the energy spectrum.

\subsection{Arrival direction studies}
\label{Sect2:anisotropy}

As described in Section~\ref{Sect2:advances}, improvements in detector
size, design and operations have led to major advances in sensitivity
for anisotropy studies.  In parallel, new Faraday rotation studies
have improved our understanding of cosmic magnetic fields,
particularly those within our Galaxy and its
halo~\cite{Pshirkov,Jansson}.  While the new observatories have not
uncovered the strong anisotropies that had been predicted by some, a
number of interesting results have ruled out several scenarios for
UHECR sources and propagation.

\subsubsection{Broad-scale anisotropy searches}
Broad-scale anisotropies are often searched for using a harmonic
analysis in right ascension (RA), though increasingly more
sophisticated multipole analyses are undertaken.  Results are
challenging to interpret, as they depend not only on distribution of
sources, but also on the distribution (including turbulence) of the
galactic and extragalactic magnetic fields, and the magnetic rigidity of the
cosmic rays.  

\begin{figure}[!t]
\centering
\begin{minipage}{.50\textwidth}
  \centering
  \includegraphics[width=1.0\linewidth]{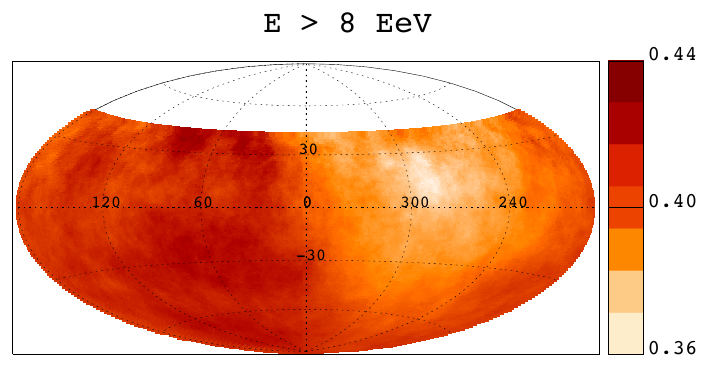}
  \caption{Auger skymap (in equatorial coordinates) for $E > 8\times
    10^{18}$\,eV.  Smoothed over windows of radius $45^\circ$, the flux
    is indicated in units of km$^{-2}$sr$^{-1}$yr$^{-1}$.  The
    significance of the implied dipole is approaching
    $5\,\sigma$~\cite{AugerAnisot_Broad2015}.}
  \label{fig:AugerAnisotropy}
\end{minipage}%
\hfill
\begin{minipage}{.47\textwidth}
  \centering
  \includegraphics[width=1.0\linewidth]{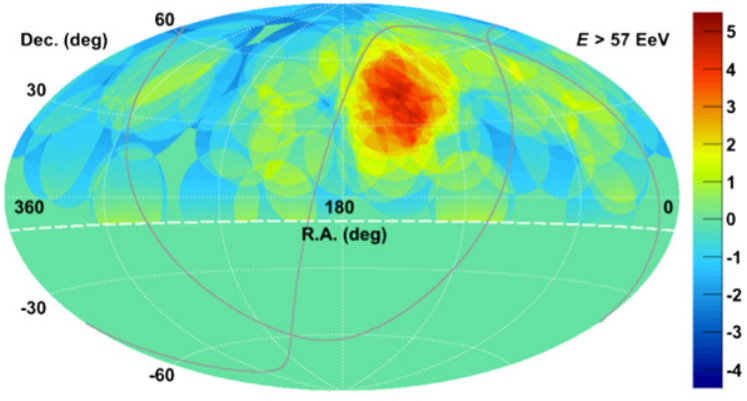}
  \caption{The TA ``hotspot'' in 2014 in equatorial coordinates.
    Nineteen events are observed above $5.7\times10^{19}$\,eV within a
    20$^\circ$ radius area of sky when 4.49 are expected, giving a
    post-trial significance of
    3.4\,$\sigma$~\cite{TA_Hotspot2014}. (Colour scale represents
    $\sigma$).}
  \label{fig:TAHotspot}
\end{minipage}
\end{figure}

One case study is the analysis by the Auger collaboration using both
the 750\,m and 1500\,m SD arrays, covering energies from around
$2\times 10^{16}$\,eV to the highest energies~\cite{SamaraiICRC15}.
For most of the reported energy range, the amplitude of the first
harmonic in RA is not significant, but the phase of the harmonic shows
an interesting energy dependence, changing from roughly the Galactic
Center direction at low energies to a direction almost $180^\circ$
away at the highest energies.  Linsley pointed out many years ago that
the phase information may have some validity even for anisotropy
amplitudes that are not significant (see~\cite{Linsley}). Two energy
bins have amplitudes approaching acceptable significance - the bin
from 1 to $2\times 10^{18}$\,eV and, especially, the bin for energies
above $8\times 10^{18}$\,eV (Figure~\ref{fig:AugerAnisotropy}).  The
latter amplitude in RA of 4.4\% has a chance probability of
$6.4\times10^{-5}$~\cite{AugerAnisot_Broad2015}.  Expressed as a
dipole amplitude and direction, the excess is $7.3\pm1.5$\%
(approaching $5\sigma$) in a direction ${\rm (RA,
  dec)}=(95^\circ\pm13^\circ , -39^\circ\pm13^\circ)$.

This apparent transition of the phase of the anisotropy from the
galactic center direction to the opposite direction coincides in
energy with the ankle of the spectrum, an energy range often seen as
the transition between galactic and extragalactic sources
(e.g.~\cite{Aloisio2012}).  It is also an energy where both Auger and
TA find that protons seem to dominate the flux (see
Section~\ref{Sect2:mass}).  The {\em dominance} of protons of
\emph{galactic} origin around $10^{18}$\,eV is excluded by the low
limits on the amplitude of the anisotropy as measured by both Auger
and TA~\cite{AugerAnisot_Broad2012,TA_Anisot2017} with TA concluding
that less than 1.3\% (95\% CL) of cosmic rays with energies between
$10^{18}$ and $3\times10^{18}$\,eV are galactic protons (given certain
assumptions about the galactic and halo magnetic fields, and assuming
an isotropic extragalactic flux).  If, in this region, an
extragalactic flux is taking over from a galactic origin, the low
level of anisotropy could be explained by the flux being the sum of
two fluxes with first-harmonic phases almost $180^\circ$ apart.

The Auger and TA collaborations have combined data to examine
broad-scale anisotropies above $10^{19}$\,eV with a full-sky
coverage~\cite{AugerTA_anis1,AugerTA_anis2}.  Only in such a full sky
analysis can a true dipole moment be measured unambiguously, and
higher moments searched for confidently.  With the current statistics,
a dipole moment of amplitude $6.5\pm1.9$\% is seen with a chance
probability of 0.5\% and a direction consistent with the Auger-only
result above $8\times10^{18}$\,eV described above.  Future joint
analyses are awaited with interest.

\subsubsection{Small and medium-scale anisotropy searches}

At the highest energies, source distances are likely to be closer than
100\,Mpc because of energy loss interactions of cosmic rays (of all
masses) on various photon fields (e.g.~\cite{Allard}).  Then, if magnetic
deflections are not too extreme, the arrival direction distribution
will mirror the distribution of sources in the local Universe.  Both
collaborations have searched for event clustering, and for
cross-correlations with various astronomical catalogs over a range of
angular scales and above a number of energy thresholds.

The most recent Auger data-set (including inclined events out to a
zenith angle of $80^\circ$) has been used for searches with energy
thresholds between $4\times10^{19}$\,eV and
$8\times10^{19}$\,eV~\cite{AugerAnisot_small2015}.  Self clustering,
and clustering around the galactic plane, the galactic center and the
super-galactic plane have been tested.  In addition, cross correlation
analyses have been performed with catalogs of extragalactic objects.
No significant anisotropies were found.  Of the studies done, the two
with the smallest post-trial probabilities (both 1.4\% as it happens)
were a correlation of cosmic ray arrival directions ($E > 5.8\times
10^{19}$\,eV) with directions of active galaxies in the Swift-BAT
X-ray catalog closer than 130\,Mpc and with luminosities greater than
$10^{44}$\,erg/s, using a $18^\circ$ search radius; and a clustering
of cosmic rays above the same energy threshold within a $15^\circ$ radius of
our closest active galaxy, Centaurus A.

The TA collaboration have done similar
searches~\cite{TA_AnisotSmall1,TA_AnisotSmall2} with similar null
results.  However, an excess on a medium angular scale of $20^\circ$ radius
has been detected above $5.7\times10^{19}$\,eV in the direction ${\rm
  (RA,dec)}=(146.7^\circ , 43.2^\circ)$.  With five years of TA data, the
``hotspot'' contained 19 events when the background expectation was
4.49~\cite{TA_Hotspot2014}.  After accounting for trials, including
the choice of angular scale, the significance of the excess is
3.4$\sigma$ (Figure~\ref{fig:TAHotspot}).  There is no obvious source
or galaxy cluster in this direction, though the excess may be
associated with large scale structure, its center being $19^\circ$ away from
the supergalactic plane.  An update of the result with an additional two
years of exposure showed a total of 24 events when the background
expectation was 6.88~\cite{TA_Tinyakov_Kyoto}.  This represents a
3.4$\sigma$ post-trial significance, no change since the original
result.  The future evolution of this analysis will be followed with
interest.


The lack of strong statistical evidence for small-scale anisotropies
at the highest energies starts to put constraints on the source
characteristics, but those constraints are tightly coupled to the mass
(charge) of the particles and the magnetic fields.  If the UHECR were
proton dominated, and if extragalactic magnetic fields are generally at
the nano-gauss scale, we would need to conclude that there was a high
source density within the 100\,Mpc horizon
(e.g. ~\cite{SourceDensity}).  On the other hand, if most of the UHECR
have medium to high charge, the lack of strong anisotropy could be
blamed on magnetic deflections.  This emphasises the importance of the
next topic in our discussion.

\subsection{Interpretations of mass composition from air shower measurements}
\label{Sect2:mass}

Unfortunately our access to information on the mass of UHECR is
rather indirect, through observations of the extensive air showers
they initiate.  We must rely on models of hadronic interactions at
extreme energies to interpret these observations in terms of the mass
of the cosmic ray.  As described in Section~\ref{Sect2:advances},
hadronic models have improved in recent years with the availability of
measurements from the LHC at centre of mass energies of up to 13\,TeV.
However, given that this corresponds to a fixed target energy of
around $10^{17}$\,eV, laboratory measurements still fall short of the
energies involved in UHECR interactions.

With this large caveat in mind, we can attempt to transform
measurements of air shower development into estimates of primary mass.

\subsubsection{Fluorescence detector measurements of shower depth of maximum} 
For many years the depth of shower maximum \xmax{} has been the prime
measurement for this purpose, first in Cherenkov light experiments
around the knee of the energy spectrum in the 1970s, and more recently
in fluorescence detector measurements at the highest energies.  \xmax{}
is the slant depth in the atmosphere (in \gcm{}) at which the air
shower reaches its maximum size (number of particles) or, near
equivalently, at which the shower reaches the maximum of its energy
deposit, $dE/dX$.  From simple arguments it can be shown that the
depth of maximum increases with the logarithm of the primary energy
for a fixed primary mass, and with the logarithm of the primary mass
number, $A$, at fixed energy (e.g. ~\cite{Matthews}).

Unprecedented resolution in \xmax{} is now possible with the
fluorescence technique, particularly due to reliable reconstruction of
the shower axis with the hybrid or stereo techniques, and partly due
to finer pixelisation and digitisation in FD cameras.  Statistical
resolution can be better than 20\,\gcm{} above
$10^{19}$\,eV~\cite{Auger_longXmax,Hanlon_UHECR2016}, with measurement
systematics below 10\,\gcm{} for Auger~\cite{Auger_longXmax} and
somewhat higher for TA~\cite{TA_Composition_Hybrid}.  Now it is
possible to confidently quote not only mean values of \xmax{} as a
function of energy, but also the width (RMS or $\sigma$) of the
distribution in some energy range.

In the interpretation of \xmax{} measurements, one needs to be aware
of any biases imposed by the detection or reconstruction processes.  A
simple example of detection bias would be a bias against the detection
of showers with very deep \xmax{} (say 900\,\gcm{}), since vertical
showers of this type would have their maxima very close to, or below,
ground level.  The Auger and TA collaborations have approached the
detection bias issue in quite different ways, both valid.  The Auger
approach~\cite{Auger_longXmax} is to apply strict cuts on the axis
geometry of air showers to avoid bias in the detection of both shallow
and deep showers.  Despite the cost of lower statistics, this allows
evaluation of the energy dependence of the ``true'' (free of detector
bias) \xmax{} distributions, which can then be compared with
theoretical predictions for various mass groups.  When computing the
RMS of the \xmax{} distributions, the experimental resolution is
subtracted in quadrature, and care is taken (with more than one
method) to account for possible undersampling of the tails of the
distributions~\cite{Auger_longXmax}.

The alternate philosophy, practised by TA~\cite{TA_Xmax2015} and
inherited from the Fly's Eye and HiRes approaches, is to only apply
cuts based on data quality, not potential bias.  The theoretical
expectation for a particular mass group is then derived using
simulations of the detection and reconstruction processes, so that any
biases and resolution effects are also present in the expectation.
This procedure maximises the event statistics for analysis.  However,
the results are not {\em easily} comparable with measurements from
other detectors.

\begin{figure}[!t]
  \centering
  \includegraphics[width=0.95\linewidth]{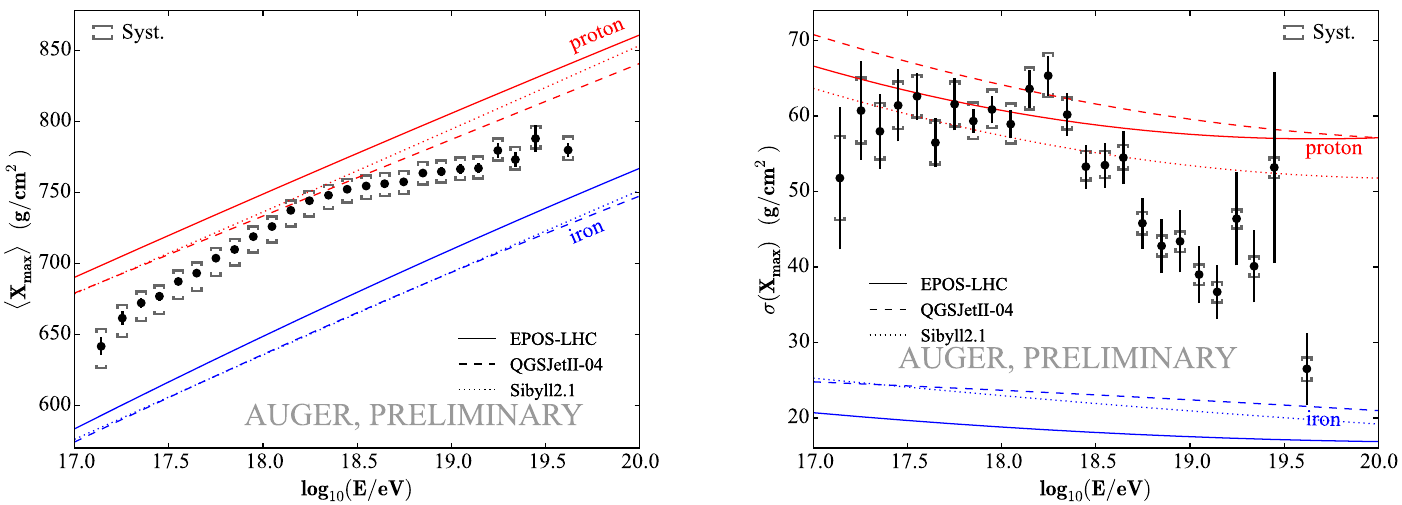}
  \caption{Auger results on the mean \xmax{} (left) and its RMS
    (right), compared with expectations for protons and iron using the
    EPOS-LHC, QGSJetII-04 and Sibyll 2.1 hadronic models.  Statistical
    and systematic error bars are indicated~\cite{AugerMass_ICRC15}.
  }
  \label{fig:AugerMass}
\end{figure}

The Auger collaboration has presented results on both the mean and RMS
of \xmax{} ($\langle X_{\rm max}\rangle$ and $\sigma(X_{\rm max})$)
from $10^{17}$ to $10^{19.6}$\,eV~\cite{AugerMass_ICRC15}, as shown in
Figure~\ref{fig:AugerMass}.  The reduction in the lower energy limit
below the previous value of $10^{17.8}$\,eV~\cite{Auger_longXmax} is
due to the inclusion of data from the HEAT FD enhancement (see
Section~\ref{Sect2:auger}).  The number of events in the latest
analysis is 23872, including 7142 events above $10^{18.2}$\,eV.  The
results can be summarised as follows,

\begin{itemize}

\item The rate of change of $\langle X_{\rm max}\rangle$ per decade of
  energy (known as the elongation rate) is not consistent at any
  energy with that expected of an unchanging mass composition, namely
  about 60\,\gcm{} per decade.  Below $10^{18.3}$\,eV the elongation
  rate is 85\,\gcm{} per decade, while above that energy it becomes
  much flatter at approximately 26\,\gcm{} per decade.  This is
  interpreted as the average mass of cosmic rays decreasing with
  energy up to the break-point, and then increasing again up to the
  highest energies.  (The lower energy elongation rate is compatible
  with the measurement by the HiRes/MIA experiments in the same energy
  range~\cite{AbuZayyad2000b}).

\item The behaviour of $\sigma(X_{\rm max})$ is broadly consistent
  with the behaviour of the mean value. Up to $10^{18.3}$\,eV the
  spread of \xmax{} is roughly constant (plausible even if the mean
  mass is decreasing, since a significant proton component appears to
  remain throughout this energy range), after which the spread appears
  to decrease with energy.

\item Using two post-LHC hadronic models, EPOS-LHC and QGSJetII-04,
  the experimental data are expressed in terms of $\langle \ln A
  \rangle$ and $\sigma^2(\ln A)$, where $A$ is the mass number of a
  cosmic ray nucleus~\cite{Auger_longXmax}.  With both models the
  mean value of $A$ is similar at the lowest energies and at the highest
  energies explored, while reducing to a minimum at around
  $10^{18.3}$\,eV.  The EPOS-LHC model interprets the data with
  slightly heavier mean $A$ at all energies, compared with
  QGSJetII-04.  At the higher energies $\sigma^2(\ln A)$ approaches
  zero for the EPOS-LHC model (implying a single type of nucleus) and
  becomes unphysically negative for the QGSJetII-04 model.

  \item \xmax{} distributions for energy bins from $10^{17.8}$\,eV to
    the highest energies have been fitted with model expectations for
    mixtures of protons with nuclei of helium, nitrogen and
    iron~\cite{AugerMassMixtures}.  With the current models, a simple
    mixture of protons and iron is not a good fit at any energy, but
    acceptable fits are obtained when intermediate masses are
    introduced.  For all models there is a significant reduction in
    the proton fraction with increasing energy above $10^{18.3}$\,eV,
    and no model requires any significant fraction of iron at any
    energy.  However, the intermediate masses concluded to be present
    at any energy have a strong model dependence.
\end{itemize}

Despite the interpretational problems associated with hadronic physics
models, the Auger results show a clear structure in the evolution with
energy of both $\langle X_{\rm max}\rangle$ and $\sigma(X_{\rm max})$.
The data do not appear consistent with a mass composition unchanging
with energy.

\begin{figure}[!t]
  \centering
  \includegraphics[width=0.48\linewidth]{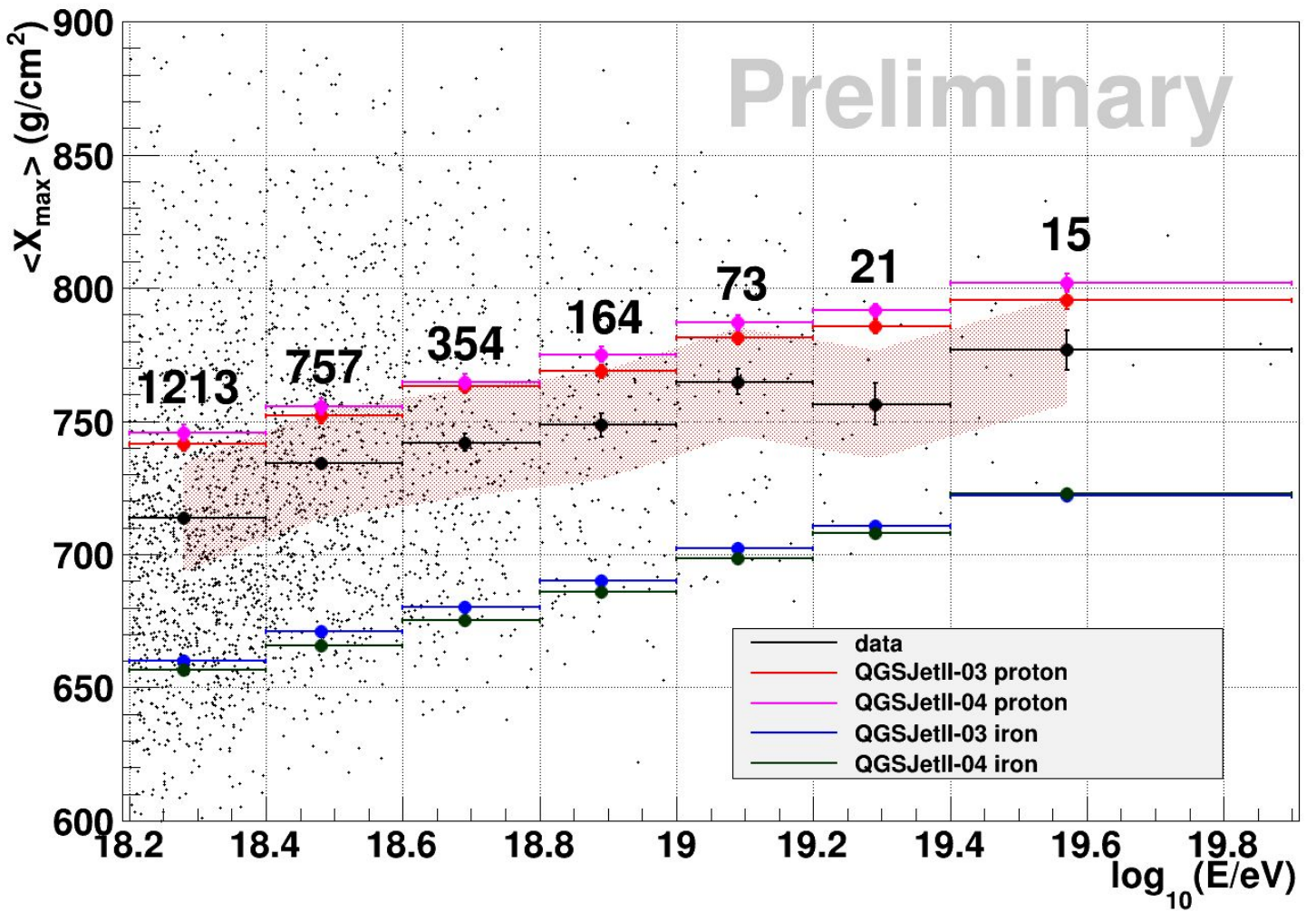}
  \caption{Telescope Array observed mean \xmax{} results from seven years of hybrid
    data from the BR/LR fluorescence detectors (preliminary data).
    Data are compared with the expectations for protons and iron from
    the QGSJetII-03 and QGSJetII-04 hadronic models.  A systematic
    uncertainty of 20\,\gcm{} is indicated by the shaded
    region~\cite{Hanlon_UHECR2016}.}
    \label{fig:TAMass}
\end{figure}

In the past three years the TA collaboration have discussed results of
three \xmax{} analyses, all using the previously discussed philosophy
of maximising statistics by applying only data quality cuts.
Detection biases are accounted for by comparing real data with
simulations having the same biases.  Those analyses are a hybrid study
of data from the Middle Drum (MD) FD detector using five
years~\cite{TA_Composition_Hybrid} and seven years~\cite{Hanlon_UHECR2016}
of exposure; a study of data from all three FDs using ``stereo''
geometrical reconstruction~\cite{TA_stereoXmaxICRC15}; and a recent
study of hybrid-reconstructed showers viewed by the Black Rock Mesa
and Long Ridge (BR/LR) fluorescence detectors over seven
years~\cite{Hanlon_UHECR2016}.  We summarise the conclusions of those
studies here,

\begin{itemize}

\item The MD hybrid study published in early
  2015~\cite{TA_Composition_Hybrid} detailed the analysis of showers
  with energies above $10^{18.2}$\,eV viewed by the refurbished HiRes
  FD detector over five years.  Using improved profile reconstruction
  cuts (based on a pattern recognition approach), \xmax{} resolution
  better than 25\,\gcm{} was achieved, with a systematic uncertainty
  in \xmax{} of better than 18\,\gcm{}.  Data were compared with
  expectations of the QGSJetII-03 hadronic model, both in terms of the
  mean \xmax{} as a function of energy, and by comparing the shapes of
  the \xmax{} distributions in a number of energy bins.  The overall
  conclusion was that, taking into account systematic uncertainties, the mean
  behaviour and the distributions are consistent with the expectations
  for a light, mainly protonic composition.  An additional two years
  of hybrid MD data were included in an analysis presented in
  2016~\cite{Hanlon_UHECR2016} with no change in conclusions.  

\item An alternative to ``hybrid'' geometrical reconstruction of FD
  events using information from the SD is to use the stereo technique
  combining views of the shower from at least two FD sites.  An
  \xmax{} analysis of stereo data from all three FD sites over seven years
  was published in 2015~\cite{TA_stereoXmaxICRC15}, with an energy
  threshold of $10^{18.4}$\,eV.  The \xmax{} resolution and systematic
  uncertainty were similar to the MD hybrid analysis above.
  Comparisons were made with 5 hadronic interaction models, including
  two making use of recent LHC input.  The trend is for more recent
  hadronic models to predict deeper developing air showers.  Based on
  the behaviour of the mean \xmax{} as a function of energy, and on
  the shape of the \xmax{} distribution for all energies, the authors
  conclude that no iron is required at any energy, and that the data
  are consistent with protons from the early QGSJet-01c model.  While
  pure protons from post-LHC models are disfavoured, a light
  composition remains consistent with the data within the systematic
  uncertainties.

\item Finally, a new analysis of seven years of hybrid data from the
  BR/LR FD stations has been presented at the UHECR 2016
  conference~\cite{Hanlon_UHECR2016}.  This data set is the largest
  with 2597 events above $10^{18.2}$\,eV, compared with 1346 events
  from the stereo analysis ($E > 10^{18.4}$\,eV) and 623 from the MD
  hybrid analysis.  With the aid of the FADC digitisation of the
  signals in these FD sites, the \xmax{} resolution is improved to
  better than 20\,\gcm{}, though the systematic uncertainty is now
  conservatively quoted as 20.3\,\gcm{}.   The data are
  compared with expectations from the QGSJetII-03 and QGSJetII-04
  models, see Figure~\ref{fig:TAMass}.  One conclusion is that,
  within the systematic uncertainty, the mean \xmax{} versus energy is
  consistent with that expected for a ``light'' composition.  In
  addition, the {\em shapes} of the \xmax{} distributions in five
  energy bins are consistent with the protonic expectations, and
  inconsistent with those of iron.

\end{itemize}

A joint group of collaborators from both Auger and TA have been
working to understand the differences in \xmax{} results from HiRes,
TA and Auger~\cite{XmaxWG2012,XmaxWG2014,XmaxWG2016}.  A particular
question is, are the differences related to experimental factors, or
due to the interpretation via hadronic models?  As we have discussed,
the comparisons are complicated by the different philosophies of the
experiments, with Auger applying cuts designed to remove detection
bias.  In their latest report~\cite{XmaxWG2016}, the group has asked
the following question: are the measurements of \xmax{} made by Auger
(both the mean values and the distributions) consistent with the
measurements of TA?  This is a question quite separate from any
particular hadronic model and mass interpretation, though such models
must be used to ``translate'' Auger measurements into TA expectations.
The Auger fractions of protons and nuclei of He, N and Fe in energy
bins above $10^{18.2}$\,eV were taken from ~\cite{AugerMassMixtures}
under the assumption of the QGSJetII-04 model.  Those mixtures were
then processed through the TA detector simulation and reconstruction
to give the \xmax{} distributions expected in each energy bin at TA
for the Auger ``mix'', taking into account any detection bias.  In
particular, the comparison was done for the seven-year, higher-statistics
BR/LR hybrid data set described above.  The conclusions are that the
Auger mix produces a mean \xmax{} as a function of energy that is
consistent with the TA measurements within the current systematic
uncertainty of 20\,\gcm{}; and that there is also qualitative agreement
between the shapes of the Auger mix distributions of \xmax{} and TA
distributions in several energy bins below $10^{19}$\,eV where TA has
sufficient statistics.  Above $10^{19}$\,eV the TA data still suffer
from insufficient statistics to come to more definite conclusions
about the distribution widths.  This important study removes much of
the doubt about the consistency of Auger and TA results, and shows the
importance of continuing dialog between the two experiments.

\subsubsection{Other mass-related measurements}
Surface detector arrays are sensitive to variations in shower
development (and hence mass) through measurements of parameters such
as the pulse rise-time in a detector, the radius of curvature of the
shower front and the lateral distribution function (e.g. see
historical examples in ~\cite{NaganoWatson2000}).  Those arrays with
particular sensitivity to muons can also attempt to tackle the mass
issue through measuring the muon content of air showers, as will be
discussed in Section~\ref{Sect2:interactions}.

Recently the Auger Observatory, in particular, has explored air shower
development with several SD methods, and we briefly mention two here.
While the resolution in inferred mass-related parameters such as
\xmax{} is typically poorer than the equivalent FD measurement, the SD
has the advantage of a 100\% duty cycle.  

The rise-time of a signal in a water-Cherenkov detector, defined
as the time taken for the signal to increase from 10\% to 50\% of the
total integrated value, is related to the core-distance of the WCD
and the zenith angle of the air shower.  It also displays azimuthal
asymmetry with respect to the azimuth of the shower axis, which can be
exploited to study shower development~\cite{AugerRiseTime2016}.  The
conclusion of the study is that above $10^{18.5}$\,eV there appears to
be an increase in the mean mass of cosmic rays, but that the detail of the
mass increase depends on the hadronic model assumed, and the core
radius range used in the analysis.  The latter dependence implies a
deficiency in both of the (post-LHC) hadronic models used.

Similar interpretational issues occur with Auger's measurements of the
muon production depth, MPD, using inclined energetic showers above
$10^{19.3}$\,eV~\cite{MPD1,MPD2}.  In such showers, the
electromagnetic component of the shower is essentially absent at
ground level, and the digitised signals from the WCDs can be analysed
to give the longitudinal profile of the production depths of muons,
and the depth of the maximum of that profile, $X^\mu_{\rm max}$.
While not the same as the depth of maximum of the overall shower
\xmax{} (dominated by the electromagnetic component), $X^\mu_{\rm
  max}$ also has sensitivity to mass.  The measurements show that the
mean $X^\mu_{\rm max}$ is effectively flat with energy above
$10^{19.3}$\,eV, implying a mass increasing with energy.  The mean
mass implied by the QGSJetII-04 model is heavy, but that implied by
the EPOS-LHC model is unphysically heavier than iron.  Again, this is
an indication that the current hadronic models are not describing the
measurements well.

Finally, there is one Auger measurement, this time at energies just
above the spectral ankle ($10^{18.5}-10^{19}$\,eV), where it is
claimed that its main conclusion is insensitive to details of hadronic
models~\cite{AugerMass_correlation}.  Here, hybrid data are used to
produce a scatter plot of \xmax{} vs $S(1000)$, and a correlation
coefficient is determined.  (The energy and zenith angle dependence of
the variables is removed before plotting).  The value of the
correlation coefficient is found to be inconsistent with any {\em
  pure} composition of {\em any} mass, with the conclusion the same
for all three post-LHC hadronic models tested.  This result disfavours,
for example, a pure protonic cosmic ray flux around the spectral
ankle, that proposed by the so-called ``dip'' model of this
feature~\cite{SpectralDip}.

\subsection{Photons and neutrinos}
\label{Sect2:photonsNeutrinos}

Apart from attempting to characterise the nuclei within the cosmic ray
flux, there is great interest in searching for photon and neutrino
candidates within the events detected by the experiments.  Photons and
neutrinos will be produced at some level in the sources due to
interactions of hadronic cosmic rays with ambient gas and photon
fields.  They will also be produced through photo-pion production when
the highest energy protons interact with photons of the cosmic
microwave background (often called ``GZK'' or cosmogenic photons and
neutrinos)~\cite{CMB1,CMB2}, and several exotic models of ``top-down''
cosmic ray production (e.g. from super-heavy dark matter) predict
significant photon fluxes (e.g. ~\cite{SuperHeavyDM}).  Thus
measurements, or limits, on the flux of UHE photons and neutrinos are
important.

\begin{figure}[!t]
\centering
\begin{minipage}{.49\textwidth}
  \centering
  \includegraphics[width=1.\linewidth]{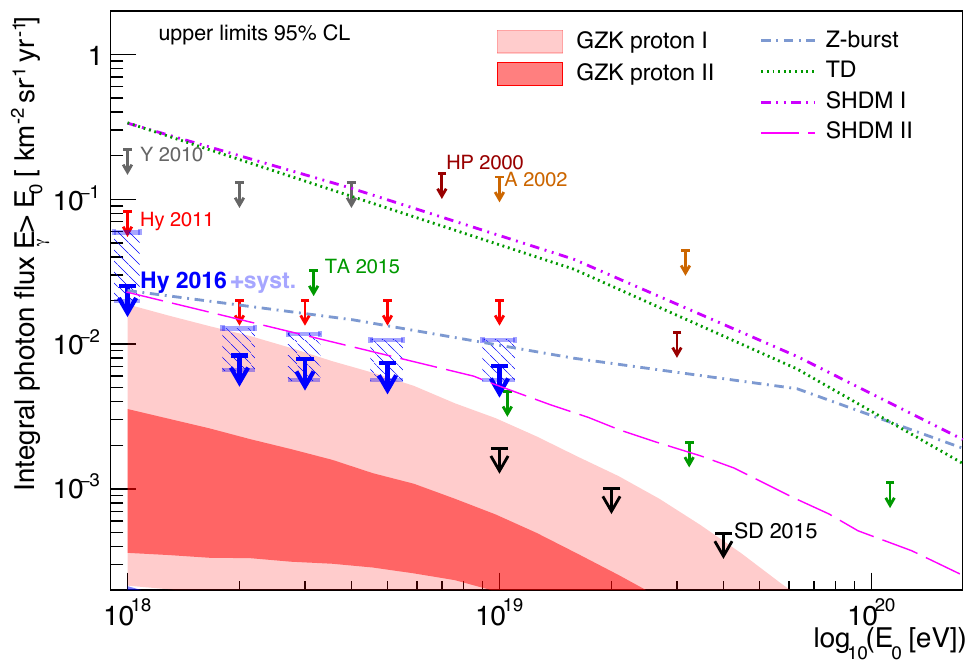}
  \caption{Diffuse photon integral flux upper limits (95\% CL) from
    Auger (black and blue points)~\cite{AugerPhoton2016} and TA (green
    points)~\cite{TAPhoton_ICRC15}.  For references to the other
    measurements and the model predictions see~\cite{AugerPhoton2016}.}
  \label{fig:Photons}
\end{minipage}%
\hfill
\begin{minipage}{.49\textwidth}
  \centering
  \includegraphics[width=1.\linewidth]{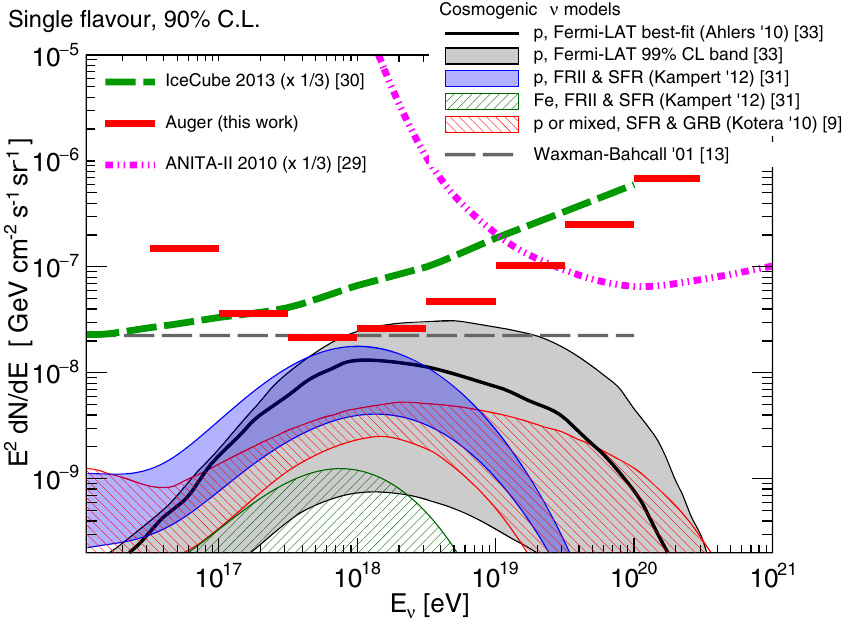}
  \caption{The Auger single-flavour limits to the UHE neutrino flux
    (90\% CL) in half-decade bins, with the equivalent limits from
    IceCube and ANITA.  For references to these other measurements,
    and the cosmogenic models shown, see~\cite{Auger_Neutrino2015}.}
  \label{fig:Neutrinos}
\end{minipage}
\end{figure}

\subsubsection{Recent photon limits} 
Both TA and Auger have produced updates to their photon limits in the
past two years.  For TA, the discrimination between hadronic and photon
initiated air showers is done using a multivariate analysis of TA SD
data using machine learning
techniques~\cite{TAPhoton_UHECR2012,TAPhoton_ICRC15}.  Among the
variables tested are a shower front curvature parameter, and the
signals in the top and bottom layers of the SD scintillators, the
latter seeking to exploit the deficit of muons in photon initiated
showers.  No photon candidates were observed with $\theta < 60^\circ$,
and 95\% CL upper limits on the integral flux were derived: 0.032,
0.0047, 0.0021 and 0.0011 km$^{-2}$\,sr$^{-1}$\,yr$^{-1}$ above 3,
10, 30 and 100\,$\times 10^{18}$\,eV respectively.

The Auger collaboration searches for photons using two techniques.  At
lower energies hybrid data are used, and a multivariate analysis of
variables including \xmax{} and a measure of the SD lateral
distribution function (LDF) is the basis of the photon
discrimination~\cite{AugerPhoton2016}.  Using 9 years of hybrid data,
three photon candidates have been identified near $10^{18}$\,eV, a number
consistent with the expected mis-classification of hadronic showers.
Thus, upper limits on the integral photon flux are calculated for five
lower energy thresholds of 1,2,3,5 and 10$\times10^{18}$\,eV, namely
0.027, 0.009, 0.008, 0.008, and 0.007
km$^{-2}$\,sr$^{-1}$\,yr$^{-1}$ (95\% CL).  This puts the photon
fraction of the flux at less than 0.1\% in the first bin and less than
2.7\% in the last.

In the decade of energy above $10^{19}$\,eV, SD data alone were used
in the Auger study~\cite{AugerPhoton_ICRC15}.  There, discriminating
shower parameters are related to the signal LDF, and the rise-times of
the WCD signals - photon showers have steeper LDFs and longer
rise-times than hadronic initiated showers.  With 8.5 years of data,
five photon candidates are observed (consistent with expectations for
hadronic mis-classification), and upper limits on the photon integral
flux are $(1.9, 1.0, 0.49) \times
10^{-3}$\,km$^{-2}$\,sr$^{-1}$\,yr$^{-1}$ (95\% CL) above thresholds
of 1, 2 and $4\times 10^{19}$\,eV.

These limits are summarised in Figure~\ref{fig:Photons}.  Note that while
the Auger results are stronger because of the larger exposure, the TA
experiment explores a different hemisphere, relevant in the case of
point sources.  The figure shows expectations for models of top-down
production of UHECR, now disfavoured at almost all energies, as are two
models of cosmogenic photons which assume a pure proton UHECR flux.
The experimental limits are encroaching on the cosmogenic model with
optimistic selections of the source spectral index and maximum energy.
The other model expectation, assuming a proton source spectral index
of $\gamma=2$ and a maximum energy of $10^{21}$\,eV, is 4 times lower
than the integral limit at $10^{19}$\,eV.  Sensitivity to this
model may be reached by the current experiments in the next decade.

\subsubsection{Recent neutrino limits}

The current competitive limits on UHE neutrinos come from the Pierre
Auger Observatory and the IceCube experiment.  The Telescope Array has
not yet published results of their searches.  IceCube and Auger have
similar sensitivities at the highest energies.

The basis of Auger's neutrino search is to identify ``young'' showers
at large zenith angles (or indeed, upward-going) in the SD dataset.  A
young shower at ground level is one with both electromagnetic and
muonic components intact.  The electromagnetic component of a large
zenith angle shower initiated by a hadron will be absorbed by the
atmosphere before hitting the ground, so ``normal'' inclined events
are characterised by SD station signals with fast rise-times and short
durations.  Auger's latest limits have combined results from three
searches to give its most sensitive single-flavour limits to
date~\cite{Auger_Neutrino2015}.  The searches include one for
earth-skimming showers (sensitive to $\nu_\tau$), and two searches in
two zenith angle bands ($\theta \in (60^\circ,75^\circ)$ and $\theta
\in (75^\circ,90^\circ)$) sensitive to all three flavours.  No
candidates were identified, and the limits are shown in
Figure~\ref{fig:Neutrinos}.  These limits are now having some
astrophysical significance, with some models of neutrino production in
sources, and exotic production mechanisms, being ruled out.  In
particular, cosmogenic neutrino production models that assume pure
proton fluxes at high-redshift sources and strong source evolution (like FR-II
galaxies) are highly disfavoured by the Auger
analysis~\cite{Auger_Neutrino2015}.  Similarly, the recent IceCube
analysis excludes with 90\% CL proton sources evolving strongly with
the evolution parameter $m > 5$ and with redshifts $z$ up to
1.4~\cite{Aartsen:2016ngq}.

In other neutrino-related studies, the Auger, TA and IceCube
experiments have reported a negative finding on a search for
coincident arrival directions of IceCube neutrinos and Auger and TA
UHECR~\cite{AugerTAIceCube}.  The Auger Observatory has also searched
for neutrino events associated with the first two LIGO gravitational
wave observations~\cite{Auger_NeutrinoGW}.

\subsection{Interaction cross-sections and tests of hadronic physics}
\label{Sect2:interactions}

As we have seen, modern cosmic ray observatories rely on models of
hadronic interactions to interpret shower development measurements in
terms of the primary cosmic ray mass.  Thankfully, the
near-calorimetric fluorescence technique has meant that energy
assignments have very little dependence on these models.  

Despite the very indirect nature of our observations of cosmic ray
interactions, modern observatories can contribute knowledge to the
nature of hadronic physics at energies well above those probed by the
LHC.  Two example areas are measurements of interaction
cross-sections, and the identification of model deficiencies in
predicting ground signals.

\subsubsection{The Proton-Air Inelastic Cross Section at Ultra-high Energies}

The first measurements of $\sigma_{\rm p-air}^{inel}$ using cosmic
rays at extreme energies were made by the
Akeno~\cite{Akeno_CrossSection} and Fly's
Eye~\cite{FlysEye_CrossSection,GaisserCrossSection} experiments,
followed later by HiRes~\cite{Belov2006}.  While Akeno showed that
this measurement was possible using a surface array (characterising
shower development using electromagnetic and muon content at ground
level), the Fly's Eye and subsequent experiments have used FD
observations of \xmax{}.  The exponential tail of a histogram of
\xmax{} measurements is fitted with a function $\exp(-X_{\rm max}/\Lambda)$ to yield the
scale of the exponential $\Lambda$.
Provided the showers contributing to the tail are initiated by
protons, $\Lambda$ can be converted to $\sigma_{\rm p-air}^{inel}$
with a relatively small sensitivity to hadronic interaction models.
For comparisons to accelerator data, the inelastic proton-air
cross-section may be converted to the inelastic and total
proton-proton cross-sections using Glauber theory (see e.g.
~\cite{Auger_CrossSection1}).

The Auger collaboration measurements were published in
2012~\cite{Auger_CrossSection1} and updated with increased statistics
in 2015~\cite{Auger_CrossSection2}.  The energy range of interest is
around $10^{18}$\,eV where the mass composition appears proton-rich.
In the latest analysis, two energy bins are used,
$10^{17.8}-10^{18}$\,eV and $10^{18}-10^{18.5}$\,eV, corresponding to
centre-of-mass energies of 39\,TeV and 56\,TeV, respectively.  Only the
deepest 20\% of the showers are used in the analysis to minimise
contamination from primaries other than protons.  Nevertheless, an
important systematic uncertainty is related to the possible
contamination by helium nuclei.  Conservatively a 25\% contamination
of helium is assumed.  Results for $\sigma_{\rm p-air}^{inel}$ are
$[457.5 \pm 17.8{\rm (stat)} ^{+19}_{-25}{\rm (syst)}]$\,mb at 39\,TeV
and $[485.8 \pm 15.8{\rm (stat)} ^{+19}_{-25}{\rm (syst)}]$\,mb at
56\,TeV.  Of the total systematic uncertainty, $\pm 10$\,mb is attributed to
hadronic model sensitivity at both energies.

The recent Telescope Array analysis is of showers observed by the
Middle Drum FD detector in hybrid mode~\cite{TA_CrossSection}. Air
showers over the energy range from $10^{18.3}-10^{19.3}$\,eV are used,
corresponding to an average centre-of-mass energy of 95\,TeV.  Showers
in the tail of the \xmax{} distribution beyond 790\,\gcm{} are assumed
to be protons.  Assuming a 25\% contamination of helium in the tail,
the $\sigma_{\rm p-air}^{inel}$ is determined to be $[567.0 \pm
  70.5{\rm (stat)} ^{+29}_{-25}{\rm (syst)}]$\,mb.

The results from both experiments have been converted to proton-proton
cross-sections to rule out the more extreme extrapolations of
accelerator data~\cite{Auger_CrossSection1,TA_CrossSection}.

\subsubsection{Characterisation of Deficiencies in Hadronic Models}
\label{Sect2:deficiencies}

Both the Auger and Telescope Array experiments have detected likely
deficiencies in the hadronic interaction models employed in air shower
simulations.  That there are deficiencies is unsurprising, given that
these models are extrapolations of direct accelerator measurements,
but it is encouraging that some of the more recent models, based on
LHC data, are less discrepant with respect to the cosmic ray
measurements (see below).

An example from the TA experiment relates to the SD energy estimator
$S(800)$.  It has been related to primary energy using simulations of
proton showers (a preference based on HiRes and TA interpretations of
mass composition) and the QGSJetII-03 hadronic
model~\cite{TA_2013Spectrum}.  The SD energy ($E_{SD}$) at a given
zenith angle is determined as the (simulation) energy that reproduces
the measured $S(800)$ at the same zenith angle.  For hybrid events,
the ratio $E_{SD}/E_{FD}$ is found to be 1.27, where $E_{FD}$, the FD
energy, is obtained calorimetrically and is essentially free of
hadronic physics uncertainties. The ratio has no significant
dependence on energy or zenith angle for $E > 10^{18.5}$\,eV and
$\theta < 45^\circ$.  The lateral distribution and other
experimentally measured variables are well reproduced by the
energy-rescaled shower simulation (proton, QGSJetII-03). The required
rescaling points to a deficiency in the simulations that predicts
fewer charged particles (electrons and/or muons) hitting the SD,
although a further quantitative analysis studying the dependence on
hadronic model, assumed mass composition and zenith angle is
necessary.  The uncertainty of the ratio $E_{SD}/E_{FD}$ is currently
dominated by the FD energy uncertainty, which is 21\% for TA, and its
improvement will help to pin down the nature and the level of the
deficiency.

The Auger collaboration has studied the muon content of inclined
(zenith angle $62^\circ-80^\circ$) air showers above $4\times
10^{18}$\,eV~\cite{Auger_Muon2015}.  At these angles the
electromagnetic component of the showers is absorbed by the
atmosphere, and the WCD signal is essentially due to muons.  Using
hybrid showers, the energy is known from the FD (to within its
systematic uncertainty of 14\%) and the muon content measured by the
SD can be compared with expectations from simulations (using two
pre-LHC and two post-LHC hadronic models) of proton, iron and mixed
compositions.  The mixed composition is that implied (for each
hadronic model) from Auger FD measurements at these energies.  A
relative integrated muon number $R_\mu$, designed to remove the energy
and zenith-angle dependence of the measurement, is used to compare
real measurements with simulations.  It is found that the simulations
underestimate the muon measurements by $(30^{+17}_{-20}{\rm
  (sys.)})$\% to $(80^{+17}_{-20}{\rm (sys.)})$\% for the assumed
mixed composition at $10^{19}$\,eV, over the range of models tested.
The models with smallest discrepancy are the post-LHC QGSJetII-04 and
EPOS-LHC models.  The quoted systematic uncertainties arise primarily
from the experimental measurement, a significant part of which is due
to the 14\% systematic uncertainty in FD energy.  In this study, the
energy systematic is necessarily transferred to a systematic in muon
number.

This mixing of systematics is largely avoided in another Auger study,
this time with more vertical ($0^\circ - 60^\circ$) hybrid showers
with energies between $6\times10^{18}$ and
$1.6\times10^{19}$\,eV~\cite{Auger_Hadron2016}. Simulated showers were
generated using the EPOS-LHC and QGSJetII-04 hadronic models for pure
protons, and for mass mixtures consistent with the Auger measurement
for each model. Then for every one of the 411 real showers, the
simulation library for a given model and mass option was searched for
the best match to the real longitudinal profile as measured by the FD.
The lateral distribution function of that simulated shower was then
compared with that measured by the SD.  On average, the simulations
underestimated the signal $S(1000)$ for both models and both
compositions, and the deficit was {\em not} constant with zenith
angle.  It is this zenith angle dependence that reduces the
degeneracy between a systematic shift in energy or muon content, since
the FD is sensitive mainly to the electromagnetic component, and the
SD is sensitive to both EM and muonic components, the mixture of which
changes with zenith angle.  The analysis results in rescale factors
$R_{\rm had}$ for the muon content and $R_E$ for the energy scale, for
each model/composition combination.  The data and the simulation can
be brought into agreement with the application of $R_{\rm had}$ and/or
$R_E$ to the simulation.  The results are that, for the mass mixture,
the energy rescale factor was consistent with unity for both hadronic
models: $R_E = 1.00 \pm 0.10$ for EPOS-LHC and $R_E = 1.00 \pm 0.14$
for QGSJetII-04, where the error is the statistical and systematic
uncertainties added in quadrature.  However, the magnitude of the rescale
factors necessary for the simulated muon numbers to match the
experiment were $R_{\rm had} = 1.33 \pm 0.16$ for EPOS-LHC and $R_{\rm
  had} = 1.61 \pm 0.21$ for QGSJetII-04, both improvements in the
significance of the model discrepancies compared with the inclined air
shower study discussed above.  Obviously, the hadronic rescale factors
required for a pure-proton composition were even larger.

Recent preliminary results from the TA collaboration indicate that the
muon content of EAS at distances between 2 and 4\,km from the core of
the shower is substantially larger by factors of two to three (at the
$3\,\sigma$ level) than predictions from any of the current hadronic
models for both proton and iron primaries.  The improved sensitivity
to muons is provided by making very selective cuts that maximise the
absorption of electrons by the atmosphere.  Results indicate that the
discrepancy increases with core distance, which may imply problems with
our understanding of the early part of EAS development, since the
muons at large core distances would originate
there~\cite{TAMuonExcess}.

These examples show that there is sensitivity for testing hadronic
models with the current observatories, taking advantage of hybrid
measurements of air showers with surface and fluorescence detectors.
We can expect even better sensitivity in the future using surface
detectors that can separately measure muon and electromagnetic shower
components, for example with the upgraded Auger
Observatory~\cite{AugerPrime}.

\section{Challenges}
\subsection{Composition}
\subsubsection{Using the \xmax{} measurement}
\label{Sect3:using_xmax}
Linsley~\cite{Linsley1977} first proposed a simple way to look for
changes in the cosmic ray composition as a function of energy. This
involves the so-called ``elongation rate'' or mean \xmax{} as a
function of energy.  In a simple superposition model, a pure single
component composition will have $\langle$\xmax{}$\rangle$ depend
logarithmically on $E$ with a constant change per decade (the
elongation rate). A change in the composition would create an energy
dependent change in this rate. A change from a light to heavier
composition would produce a decrease in the rate of change of mean
\xmax{} with energy (a flatter or even negative elongation rate) for
example. As long as there are no rapid changes in hadronic interaction
physics, this is true in a model independent way. However, the
elongation rate does not tell us what the composition actually is. For
this, hadronic models must be used to simulate air showers, the
response of the fluorescence detector must be folded in or dealt
with using cuts, and the resultant absolute position of
$\langle$\xmax{}$\rangle$ compared with data at a number of energies.

There are a number of problems with this approach. Firstly, the
absolute predicted value of $\langle$\xmax{}$\rangle$ is hadronic
model dependent, with variations of 10-20\,\gcm{} between extreme
models at any given energy. Then, the actual \xmax{} distribution is
asymmetric and if there is a significant protonic component it will
have a long tail extending to deep \xmax{}. Heavier nuclei will have
less pronounced tails. The mean value, $\langle$\xmax{}$\rangle$, is
sensitive to these tails which can be affected by detector
systematics. This is one possible source of bias that can produce a
systematic difference between simulations and data.  Studies indicate
that all the various effects can produce a net residual systematic in
the mean \xmax{} as large as 10-20\,\gcm{} for the TA
experiment~\cite{TA_Composition_Hybrid}, and up to 10\,\gcm{} for the
Auger experiment~\cite{Auger_longXmax}.  Unless care is taken,
undersampling due to low statistics may also shift
$\langle$\xmax{}$\rangle$.  The second moment of the distribution, the
RMS, is sensitive to both the tails and the width of the distribution
and hence carries additional information. The RMS is less hadronic
model dependent since the distribution width mostly depends on
superposition. A change of RMS from 60\,\gcm{} (characteristic of
proton showers for essentially all hadronic models) to 30\,\gcm{} as a
function of energy is observed by the Auger
collaboration~\cite{Auger_longXmax} in the energy range above
10$^{18}$\,eV and it can be considered evidence for a change in
composition. The smaller RMS can only be produced by heavier nuclei
such as CNO or Fe, again in an essentially model independent way.
However, the RMS measurement suffers some of the same systematic
problems as the elongation rate.  Undersampling of the tail of a
distribution either due to low statistics or detector bias can mimic a
composition change.  The Auger collaboration has been able to address
this in some detail because of its high
statistics~\cite{Auger_longXmax}. The TA measurements at the highest
energies still suffer from insufficient statistics to address this
issue completely.

A puzzling issue that has emerged from this approach is 
that it is difficult to reconcile the $\langle$\xmax{}$\rangle$ with the RMS 
distributions. In a simple two component p/Fe model for example, 
the RMS at the highest energies agrees well with a nearly pure Fe composition but 
the $\langle$\xmax{}$\rangle$  requires a much lighter mix.  Reproducing this is 
a struggle even with a four component composition.

For all these reasons, comparison of the full \xmax{} distribution 
between data and Monte Carlo (MC) simulations seems the best approach. In principle 
it provides the maximum information.  However, a straightforward 
statistical comparison of data and simulations is  made impractical 
because of the presence of significant systematic uncertainties both 
in the data (overall \xmax{} position) and the hadronic model.

There are two approaches to deal with the problem of determining
composition. The TA and HiRes collaborations apply loose cuts to data
(sufficient to ensure good resolution) and carefully simulate p, He,
N and Fe air showers based on a variety of hadronic models. Whatever
distortions in the \xmax{} distribution are generated by the detection
method and reconstruction should then be evident in the reconstructed
simulated data. The Auger collaboration instead applies much tighter
fiducial volume cuts which minimise any detector and reconstruction
bias. The resulting data can then be directly compared to the
``thrown'' simulations. Direct comparison of the data from these two
approaches can be problematic since the detector distortions will be
different, though the biases in the most recent TA hybrid analysis are
much smaller than for previous results.

Recently, the Auger and TA groups have developed a method to improve
comparison of \xmax{} distributions~\cite{XmaxWG2016}. The Auger group
fits their cut and unbiased data to a simulated composition mixture as
a function of energy. The resultant composition fractions are then
used by the TA group to generate ``thrown'' simulations.  These are
then processed through the TA reconstruction process and compared to the
data. Preliminary results show agreement within the systematic uncertainty for
the overall elongation rate. This approach is, in principle, independent
of the hadronic model used, since this is only used as a method to
port one data set into another experiment's acceptance.

But how does one deal with systematic uncertainties in these comparisons?
What is needed is a comparison method that allows for a sliding
\xmax{} scale (to take account of overall systematics) while
preserving the shape of the distribution. The TA group has proposed
such a method~\cite{Hanlon_UHECR2016} which first removes the energy
dependence of the distributions and allows an \xmax{} shift for the
data which is determined by the best overall \xmax{} profile fit. One
can then compare the distribution shape goodness of fit to the
required sliding \xmax{} scale factor to see how well any given
composition assumption does when compared to the data. For example, if
the scale factor shift required is well beyond the estimated
systematic uncertainties and there is a poor profile shape fit, then that
hypothesis can be discarded.

Another approach, used by Auger~\cite{AugerMassMixtures}, is to
directly compare a multi-component mix of simulated showers with the
data. This is done for a variety of hadronic models. This approach
uses p, He, N and Fe as markers for the actual cosmic ray
composition. The QGSJetII-04, Sibyll 2.1, and EPOS-LHC hadronic models
are used.  The individual components have separate weights that vary
as a function of energy and the experimental systematics are folded in
to the degree that they are known. The results change as one shifts
hadronic model assumptions; while an overall trend of moving from p
to He in the energy range from $10^{18.3}-10^{18.8}$\,eV is shared by
all 3 models, only EPOS-LHC gives a nearly constant and large fraction of 
 N (~40\%) across the full energy span, while the other two models are consistent
with almost zero N fraction.  Above $10^{18.8}$\,eV, dominant components are He
and N, but their proportions are very much model dependent; EPOS-LHC favors
N while the other two models strongly support He.  These differences make an
astrophysical interpretation challenging as they are most likely due
to model inadequacies.  While the details are not clear, the required
proton fraction decreases above 3$\times$10$^{18}$\,eV and there is no
requirement for any significant iron fraction.

The lack of consistency is not surprising given that even for a single
hadronic model, HiRes publications~\cite{Abbasi2005b} have noted that
introducing more than two components into a fit to an \xmax{}
distribution does not lead to an easily interpretable result as
various combinations can give equally good fits. In the case of Auger,
the best fits are produced with more than two components, but the
uniqueness of the interpretation remains problematic.

Any particular approach to reconstruction shower profiles has hidden
systematics which are intrinsic to the chosen approach and the
particular software implementation. This systematic uncertainty is separate
from detector or atmospheric systematics. Two different, error-free,
reconstruction programs that use different approaches (different
binning, least-square fitting routines, tabular vs functional
corrections etc.) will produce slightly different results. The TA
group has explored this ``intrinsic'' systematic by comparing
completely independent and otherwise well-vetted hybrid reconstruction
programs as well as by comparing results from stereo data. They find
that it is very difficult to make the \xmax{} distributions (for the
same data or simulations) agree to better than
5-10\,\gcm{}~\cite{HanlonPC, Auger_longXmax}.  This seems to be an
irreducible systematic uncertainty.

A particular complication in the study of cosmic ray composition is
the fact that any nucleus heavier than a proton will eventually
fragment to a lighter nucleus as it travels from its source to the
Earth. This fragmentation is due to the interaction of the nucleus
with both the relic 2.7\,K black body photons and the IR radiation
fields produced by stellar
radiation~\cite{Puget1976,Stecker1999,Aab2016,Allard}. As a result,
even a pure single nucleus composition heavier than a proton at the
source should appear as a mixed composition.  A pure proton primary
composition will arrive intact, but observation of a proton component
cannot rule out that part of this component is due to heavier
nuclei. On the other hand, observation of an iron component uniquely
indicates the existence of a primary iron at the source, since stellar
nucleo-synthesis does not provide any significant concentration of
nuclei above iron.  Propagation models show that He, CNO and Fe have
different spallation probabilities as a function of
energy~\cite{Taylor2006,Hooper2008}.  This is particularly evident
above 3$\times$10$^{19}$\,eV where the He mean free path is on the order
of 10 Mpc, compared to $\sim$ 100 Mpc for N~\cite{Allard}.  A He
dominated flux above 3$\times$10$^{19}$\,eV only makes sense if sources
are very close. The lack of anisotropy makes any such assertion
implausible. Indeed propagation calculations indicate that integrated
over the large scale structure, the mean $A$ of nuclei originating as
He is essentially one~\cite{Hooper2008}.  If the observed He is the
result of fragmentation of heavier nuclei, then a proper proportion of
these heavy nuclei, whose mean free paths are much longer, must also
be seen in the composition distribution.  Incorporating the observed
cosmic ray nuclear abundances found with a particular hadronic model
directly with propagation effect weights in a more direct fashion than
is currently done could be very helpful. In some cases, this may rule
out an otherwise well fitting model as leading to astrophysically
implausible scenarios.

\subsubsection{Implications of the lack of iron in UHECR}
\label{Sect3:lack_of_iron}
While the systematic uncertainties in \xmax{} determination and comparison to
simulated compositions are still too large to make strong statements
about the relative abundance of elements in the cosmic ray flux at
Earth or at their origin, we have learned that there is very little
iron in the flux above 10$^{18}$\,eV.  Given existing systematic uncertainties,
it is safe to say that any direct or secondary heavy nuclei from Fe to
Si are absent from the spectrum. We know this absence with better
precision than we know what elements are present in the flux. What
does this imply about the sources and acceleration mechanisms of
UHECR?  UHE primary iron can easily reach the Earth from as far
away as 100 Mpc and its spallated byproducts down to Si from much
further distances. Are magnetic field effects strong enough to
substantially increase the effective path length? Is there a
deficit of iron in the cosmic material feeding the accelerator? There
is astronomical evidence that iron attaches itself to dust particles
and hence appears to be somewhat depleted in its free form, for
example~\cite{DeCia2016,Krogager2016,Gail2015}.  However, why the
iron-rich dust particles cannot be swept from an accretion disk into
the accelerator beam, decomposed to their atomic constituents and
provide the original iron abundance is not clear.  With a charge 26
times that of a proton, iron will be accelerated quite efficiently at
the highest energies.  If iron is indeed accelerated at the source
then its absence must require photon fluxes at the source that
essentially eliminate it from the cosmic ray flux.  The absence of
heavy elements in the cosmic ray spectrum may thus be an important
constraint and clue to cosmic ray origins.

\subsubsection{FD/SD energy mismatch, muon excess}
\label{Sect3:missmatch}
Much of the progress in establishing the structures in the cosmic ray
spectrum come from the reliable energy scale provided by air
fluorescence. With a $\sim$15\% FD energy resolution and similar
systematics, calibrating the SD energy scale to simultaneously
observed FD events has made the SD spectrum energy largely hadronic
model independent~\cite{Song2000}. For the TA case, the energy scale
adjustment that needs to be made between the FD and SD is on the order
of 25-30\% if one uses QGSJetII-03 proton simulations for the SD
energy.  The measured particle densities produce a lateral
distribution which, were it analyzed based on hadronic model
simulations of air showers, would generate too high an energy by this
amount. In other words, there are too many charged particles at ground
level for a shower with an energy as determined by the FDs.  Because
Auger water Cherenkov detectors are quite sensitive to muons, this
mismatch has been attributed to an excess of muons in the
data compared with expectation.  In the case of TA's plastic scintillation detectors, electrons
and muons have similar detector response and the mismatch can only be
partly attributed to a muon excess.  Studies are proceeding to
investigate whether scintillation detectors at large distances from
the core which should have mainly muon initiated signals are
consistent with the Auger results~\cite{TAMuonExcess}. In any case, it is clear
that the hadronic models that are used to simulate showers are not
adequate. Until this issue is resolved it is difficult to use muon
density to measure cosmic ray composition precisely, though trends can
certainly be established (see below).

\subsubsection{Other techniques}

\paragraph{a. Radio} 
The fluorescence technique revolutionised the study of UHECR physics
because it made possible a largely calorimetric determination of the
energy scale and a relatively direct measure of composition using
\xmax{} of the showers. It requires clear moonless nights which
restricts its on-time to 10-15\% of the SD operation times.  Recently
a great deal of work has been done in investigating the possibility of
using radio emission from EAS in much the same way as one now uses
fluorescence~\cite{huege-kyoto}.  Radio can, in principle, determine
the shower energy and \xmax{} and would have the advantage of
$\sim$100\% on-time. A number of radio arrays have now been operating
either stand-alone or in conjunction with surface and air fluorescence
detectors~\cite{lofar_xmax, aera_instrument, aera_observation}.
Because there are several mechanisms in the air shower development
that can generate radio waves, the detailed simulation has taken some
time to develop but now seems sufficiently advanced. Meaningful
comparisons with real data have been done and good agreement is now
evident between simulations and radio and SD
measurements~\cite{radio-1}.  These studies have been largely limited
to energies less than 10$^{18}$\,eV however, and in this energy regime it
appears that an array of radio antennas with spacings not dissimilar
to SD spacings are required for good energy and \xmax{} resolution.
If similar spacings is required for $>$ 10$^{18}$\,eV energies, the
costs associated with instrumenting $>$ 1000 km$^2$ arrays become
significant. Until more complete optimisation and cost/benefit
analyses for the UHECR regime are done it is not clear that this
technique will supplant fluorescence and particle SD arrays. In any
case, significant physics from the low energy arrays is required
before this new technique can be considered fully vetted.

\paragraph{b. \xmax{} - SD signal correlations}
Recently an approach to studying composition has been proposed using
the correlation between \xmax{} and the SD
signal~\cite{AugerMass_correlation}. This is based on the very old
idea that if iron and proton showers have different \xmax{} distributions and
different $N_\mu$ distributions then superposing their \xmax{}-$N_\mu$
scatter plots will lead to a negative correlation even though the pure
distributions have a positive correlation. Since this is generally
true of any hadronic model, the claim of this approach is that this is
more model independent than either a pure \xmax{} or pure $N_\mu$
analysis. However, since neither Auger or TA actually measure $N_\mu$
(except at large zenith angles in the case of Auger) the searched for
correlation is with $S(1000)$ or $S(800)$. The recent work on this by
Auger~\cite{AugerMass_correlation} shows that in the
$10^{18.5}-10^{19.0}$\,eV energy region the correlation is
inconsistent with a pure composition. A preliminary study by the TA
collaboration on the other hand shows no inconsistency with the
assumption of a protonic composition~\cite{LindquistPC}. However, in
the case of TA, the method is not nearly as sensitive as the \xmax{}
method. This may be because the muon content in TA is not as large a
component of the SD signal as for Auger.  Since the method is ``de
facto'' dependent on detector muon sensitivity it must also be to some
extent model dependent, although the Auger study checks this with two
independent models.  Studying the applicability of this method in the
lower energy region ($10^{18}-10^{18.5}$\,eV) would be of interest
since there the composition is likely to be more pure, given both the
Auger and TA \xmax{} data.

\subsection{Energy}
\label{Sect3:energy} 
One of the most significant results coming from Auger and TA is the
overall agreement in the shape of the UHECR spectrum. At first glance,
both spectra show a clear ankle structure and a cutoff, although the
precise energies for these structures differ.  However, a shift of
either experiment's energy scale by 10-15\% brings the ankle structure
into excellent agreement~\cite{Verzi_PTEP}. Since such a shift is within the
systematic uncertainties of either experiment, it would seem that there are
no significant north - south differences here. A closer look at the ratio
of the two spectra shows, however, that the location and shape of the
cutoff seems different at the $\sim$3$\sigma$
level~\cite{VerziICRC15,TA_Spectrum_ICRC15}.  The Auger-TA combined
working groups have looked at this difference and, so far, have found
no reason to believe it is a result of systematic uncertainties in energy. If
this is truly a difference in the flux of northern and southern
sources at the highest energies, there should be an overall
declination dependence. Preliminary evidence from TA indicates that
the TA spectrum becomes much more like the Auger spectrum near the
cutoff if a declination cut of $<25^\circ$ is
made~\cite{IvanovPC}. The difference (a higher energy cutoff for
TA) must then come from higher declinations which also contain the
``hot spot'' that may be a signature for a relatively nearby source.
While this is suggestive, much more work needs to be done to
demonstrate that this cannot be a systematic effect either in energy
or aperture estimation.

\subsubsection{Energy Scale Shift Systematics}
As indicated above, the ankle structure which is seen with high statistics 
in both TA and Auger data can be used to estimate the difference in energy 
scale of the two experiments. While the result is within systematic uncertainty 
estimates, it is important to understand the nature of the energy shift 
as well as possible. Given the current precise nature of shower 
reconstruction, differences in energy can most likely be attributed to 
systematic uncertainties in optical properties (mirror reflectivity, light 
collection efficiency etc.), phototube gain calibration, atmospheric 
transmission and air fluorescence efficiency. All but the last are by 
their nature detector dependent and we must rely on the diligence of 
the experimenters in estimating how well they know these parameters.

The air fluorescence efficiency is in principle a common factor,
though it depends on humidity and temperature corrections which may be
somewhat different in the two locations. For reasons of keeping a
historically consistent energy scale, the HiRes and TA groups have
used the original Kakimoto et al. overall yield
measurement~\cite{Kakimoto} while a subset of Auger collaborators has
launched a series of special experiments to measure the fluorescence
yield more precisely with AIRFLY~\cite{AIRFLY0,AIRFLY}. The HiRes
group also performed a series of measurements using an electron beam
at SLAC (FLASH~\cite{FLASH2, FLASH3, FLASH4}) but only the relative
spectral line strength measurement has so far been incorporated in the
TA analysis. The TA experiment also includes a 40 MeV electron linac
whose vertical beam is seen in the field of view of one of the
fluorescence detectors. Work on understanding the results of this
in-situ measurement is proceeding~\cite{TA_ELS_ICRC15}. All
contemporary measurements of the absolute value of air fluorescence
rely on a fixed energy electron or proton beam which deposits energy
in a small pressure controlled chamber.  Significant corrections for
deposited energy escaping the chamber (in the form of delta and gamma
rays) must be made. The MACFLY~\cite{MACFLY2007} and thick target
FLASH experiments generated a shower in an air-equivalent material and
observed the fluorescence as a function of absorber. Neither of these
experiments was able to produce an absolute value for the
fluorescence efficiency with sufficiently small uncertainties to compete with
the thin target experiments, though they did show that the relative
longitudinal development of showers is well tracked by the resultant
air fluorescence.  Recently a new experiment at SLAC called
sFLASH~\cite{SokolskyPC} is attempting a $<$ 10\% total systematic
uncertainty measurement of air fluorescence from a $\sim$10 GeV electron
shower developing at sea level and observed near shower maximum.  If
successful, this will be a valuable cross check on the thin target
results. What is lacking is a common air fluorescence result that is
used for both TA and Auger analysis. It is to be hoped that such a
convergence can occur in the near future.

\subsection{Anisotropy}
\label{Sect3:anisotropy}
Large scale anisotropy can be searched for using a multipole
expansion.  This is however tricky to do without bias unless one has
full coverage over the celestial sphere. It is thus very advantageous
to combine TA and Auger arrival direction data. There are several
challenges to using this data set however.  Because of the energy
scale difference one cannot simply apply the same energy cut for both
data samples. There are also potential systematic differences in
determining the detector apertures of the two detectors. The Auger/TA
anisotropy working group has developed an approach that uses the
overlapping declination band for the two
detectors~\cite{AugerTA_anis1}. The fluxes in this band are normalised
and this normalisation is carried over to the total data set.  The
assumption here is that the spectrum has no significant declination
dependence in the overlap band.  The resultant distributions have yet
to show any statistically significant dipole or quadrupole moments,
though Auger itself observes a significant dipole
enhancement~\cite{AugerAnisot_Broad2015}.  A better understanding of
the energy scale shift between the two experiments, and strategies to deal
with it, could provide a simpler method of combining data without
additional assumptions.

\subsection{TA hot spot} 
\label{Sect3:hotspot}
With the fading of the Auger association of UHECR with
AGN~\cite{AugerAGN2007,AugerAnisot_small2015}, the community's hope
for finding clear associations of cosmic ray arrival directions with
astrophysical sources has received a lift with the possible observation
(at the 3.4$\sigma$ level) of a concentration of cosmic rays with
energies above 5.7$\times$10$^{19}$\,eV in the northern sky by the TA
experiment~\cite{TA_Hotspot2014}.  This ``hot spot'' of $20^\circ$
radius is observed near Ursa Major, about 10 degrees off the
supergalactic plane.  If this intermediate-scale anisotropy is
confirmed with more statistics its location raises interesting
questions, since none of the previously assumed cosmic ray sources
(e.g. the Virgo cluster) are in the immediate vicinity. If the sources
are actually in the adjacent portion of the supergalactic plane, then
there must be a magnetic field effect shifting the flux to the
observed location. A suggestion has been proposed that there is a
magnetic flux tube produced by a filament of galaxies connecting the
hot spot to sources such as M87~\cite{Ryu2016}.  Another possibility
is M82 which is sufficiently close to account for the hot spot using
currently estimated magnetic fields.  Tidal disruption events creating
one or more flashes of extremely high energy protons or nuclei have
been proposed for the acceleration mechanism~\cite{Pfeffer, He}.  If
this hot spot strengthens in significance it will pose a challenge to
our understanding of sources and magnetic field configurations.

If medium or small-scale anisotropy is finally observed the next major 
challenge is to correlate our composition related  information with 
arrival direction information. Hybrid FD plus SD data would be the most 
convincing, but requires the most running time for any given source. If the 
muon content of showers can be better understood and correlated with 
composition, this could give the most sensitive composition dependent 
anisotropy measurement. It is unfortunate that the currently most 
likely source (TA hot spot) and the major Auger upgrade of their SDs 
to better detect muons correspond to disconnected parts of the sky. 
If the hot spot is confirmed, and the muon content becomes better 
understood, coming to grips with this issue will be one of the major 
challenges for this community.

\section{Future Observations}

\subsection{Extension and Upgrade of Ground Observatories}

Above the ``knee'' at around $10^{16}$\,eV, the cosmic ray energy
spectrum and \xmax{} measurement demonstrate rich features, and around
$10^{19}$\,eV and above, various anisotropies seem to show up in the
energy spectrum and flux. Where statistics are adequate, no obvious
inconsistency is found in the \xmax{} measurements in the northern and
southern hemispheres above $10^{18}$\,eV, but their interpretation
allows a range of composition mixes and energy dependencies due to
statistical and systematic limitations. It is important for Auger and
TA to cover the entire sky and the whole energy region together, in
order to bring these indications to a consistent set of observational
facts.  It will become the basis of locating the galactic to
extra-galactic transition energy of cosmic rays sources, and of
building a viable astrophysical model to explain the production and
propagation of UHECR.  Continuing to challenge this physics, TA and
Auger are both planning to start the operation of extended and upgraded
detectors around 2018--19.

In the northern hemisphere, TA$\times$4, the extension of
TA~\cite{sagawa_icrc2015, kido_uhecr2016} is in preparation.  It will
extend the aperture of the SD by a factor of four by 2018. Leaving a
part of the SD intact with 1.2 km spacing, an extended part will have
a 2.08 km spacing, together covering a 3,000 km$^2$ ground
area. Adding two more FD stations, the hybrid coverage will be
tripled.  The trigger efficiency of the extended SD will be larger
than 95 \% for E $>~10^{19.8}$\,eV.  Resolutions will be slightly
compromised to become $\sim$25 \% for energy and 2.2$^\circ$ for the
arrival direction.  In three years of running over 2018-2021, the number
of SD events above 57\,EeV (=$10^{19.76}$\,eV) will be quadrupled to
become 300, of which $\sim$80 would be in the hotspot region, assuming
the flux of \cite{TA_Hotspot2014}.  The measurement range of
$\langle$\xmax{}$\rangle$ using hybrid events will be extended to
$\sim 10^{19.6}$\,eV from the present
$10^{19.4}$\,eV~\cite{Hanlon_UHECR2016}.  The SD design of the
TA$\times$4 was re-optimised to use a much shorter length of
wavelength shifting fibers (1/3 of the length in the TA/SD) while
keeping the same number of photo-electrons collected by the PMT for a
minimum ionising particle.  The quantum efficiency and the linear
range of the PMT is nearly doubled.

In the southern hemisphere, the Auger Observatory plans to upgrade the
detector to AugerPrime by 2018~\cite{AugerPrime, engel_uhecr2016}.
All the $\sim$1600 stations will be equipped with a 3.8 m$^2$ plastic
scintillator on top and the waveform sampling electronics will become
three times faster (to 120 MHz).  An integrated analysis of water-Cherenkov
and scintillator signals will enable an isolation of muonic and
electromagnetic (EM) energy deposits, and enable the counting of the
number of muons hitting the SD.  New methods are being developed to
estimate \xmax{} from SD measurements alone, taking advantage of
so-called shower universality~\cite{Universality}.  The search for
small-scale anisotropy and source correlation is expected to improve
significantly by selecting SD events with high likelihood of being
protons or light nuclei.  The muon identification is double-checked
for a portion of SD events using an array of scintillators buried 2.3
m underground.  The duty cycle of FD operation is expected to become
1.5 times larger by tolerating data collection with a higher night sky
background.  The mixed composition result~\cite{AugerMassMixtures}
will be further checked with measurements from the FD together with
the enhanced SD with its own measurements of \xmax{} and muon content.
AugerPrime and TA$\times$4 together will have all-sky coverage with a
total of 6,000 km$^2$ of surface area; one at 39$^\circ$ North and the
other at 35$^\circ$ South.  The overlapping region at low declinations
($-16^\circ$ to $+45^\circ$) will be important in understanding the
relative exposures and to examine systematics of the detectors and
data analyses.

Air shower detectors operating in the last decade have reported a
series of Earth-science related findings; TA's SD recorded bursts of
particle showers associated with lightning~\cite{okuda_icrc2015,
  belz_uhecr2016}, the development of distant atmospheric ``elves''
was recorded by Auger~\cite{tonachini_icrc2013,
  colalillo_uhecr2016}, and the LOPES radio signal from air showers was
modulated by thunder-clouds~\cite{buitink_thundercloud} etc.. UHECR
observatories may become an interesting research tool for Earth and
atmospheric sciences in the next decade.

\subsection{Development of Radio Detection}
Understanding the mechanism of air showers generating radio signals in
the sub-100 MHz range has advanced greatly in the last decade (see
\cite{huege_pr, huege_uhecr2016} for reviews).  Newly developed
simulation codes tell us that the radio signal comes from two types of
time-varying, fast-moving effective charges generated in the air
shower; one is the lateral movement of shower e$^\pm$ under the
geomagnetic field and the other is the longitudinal movement of net
charge in the shower front (the Askaryan effect).  Both signals scale
with the square of the electromagnetic energy ($\propto E^2$).  The
signal is sharply forward peaked in the direction of air shower
development and stands well above the galactic radio noise for
energies exceeding 10$^{16}$\,eV.  The radio telescope LOFAR, operating
in cosmic ray detection mode, realised a very fine radio sampling of
air showers, and succeeded in observing air showers of energy
10$^{17}$ - 10$^{17.5}$\,eV with a typical \xmax{} reconstruction
uncertainty of 17\,g/cm$^2$~\cite{lofar_xmax}.

The AERA radio array with varied antenna spacing has been deployed at
the Auger site for testing the detection of the highest energy
showers~\cite{aera_instrument,aera_observation}.  The results
demonstrate that a dense deployment of antennae is required for the
effective detection of UHECR that have a footprint of several 100
meters in diameter.  Even though the elements of the RD (Radio
Detector) may be relatively simple and inexpensive, the total cost of
deploying and operating a large area detector would become
prohibitive. One practical application for UHECR is for the
measurement of very inclined air showers with $\theta > 70^\circ$,
which has an extended oval footprint larger than 10 km$^2$.  Such RDs
deployed together with the SD may also be used for the calibration of
SD energy, making use of the fact that the radio signal originates
predominantly from the EM component of the shower, and that it has a
negligible attenuation in the atmosphere. Note that this is the
kinematic region where precise measurements by standard SD techniques,
using a water-Cherenkov station or a scintillator, have large uncertainties,
and redundant information is useful.  The radio detector is also
expected to have a high duty factor of $\sim$95\%.

Another direction of progress foreseen for using radio signals from
EAS is the detection of high energy neutrino-induced showers in the
Antarctic ice via the Askaryan effect. Pioneering work searching for
such short, GHz polarised radio signals from the horizon in Antarctica
began with the ANITA balloon experiment in 2006~\cite{anita_1}.  Its
4th flight was launched in December 2016.  The ARA and ARIANNA
experiments were recently proposed and extensive RD is underway to
detect the Askaryan signal from cosmogenic neutrinos near or on the
surface of the Antarctic ice~\cite{barwick_ana_arianna}.

Searches for GHz ``Molecular Bremsstrahlung'' radio emission from
particle showers in the atmosphere~\cite{mol-brems_gorham} have so far
not been successful~\cite{mol-brems_crome, mol-brems_els}.  Also, a
limit has been set by the TARA experiment at the TA site for detecting
the modulation of 54.1\,MHz carrier radio waves by the ionised column
generated by a UHECR shower in the atmosphere~\cite{radio-5}.

\subsection{Observations from Space}
The EUSO international collaboration was formed in 2000 to install a
wide field of view (FoV) telescope at the International Space Station
(ISS) to look down on the Earth's atmosphere and search for air
fluorescence flashes from UHECR~\cite{euso_proposal}.  The JEM-EUSO detector
employs a Fresnel lens telescope with a diameter of 2.4 m and a
60$^\circ$ FoV, covering a ground area of 200\,km radius from an
altitude of 400\,km~\cite{jemeuso_acceptance}.  The effective ground
coverage with the expected duty factor of 20\% is 28,000km$^2$, or
approximately five times that of AugerPrime and TA$\times$4 combined. A
tilted mode of observation would increase the acceptance by a factor
of three or more at the cost of reduced resolution and higher detection
threshold.  The ISS inclination angle of 51.6$^\circ$ allows a
uniform survey of the Earth's atmosphere in the northern and southern
hemispheres with nearly the same acceptance and event geometry.
The observations from high altitude and the limited optical entrance pupil
of the Fresnel lens will however limit the JEM-EUSO detection
threshold to be $\sim 10^{19.5}$\,eV, and the \xmax{} resolution is foreseen
to be larger than 60\,g/cm$^2$ in the nadir
mode~\cite{jemeuso_resolution}, making the differentiation of nuclear
composition difficult.

The mission schedule of JEM-EUSO is yet to be determined, but an
extensive series of tests of the prototype instruments is being
performed~\cite{casolino_uhecr2016}. Major efforts include the balloon
borne EUSO-SPB test (2017), and deployments of mini-EUSO (2017) and
K-EUSO (2020) at the ISS.  The K-EUSO experiment will have a segmented
Fresnel mirror 3.4 m in diameter, and its effective coverage of the
ground will be 6,200 km$^2$ above 10$^{19.5}$\,eV, about equivalent to
AugerPrime and TA$\times$4 combined.  An exploring Russian satellite
experiment, TUS, with similar optics was launched in 2016 and is being
commissioned~\cite{tus_uhecr2016}.  The uniform all-sky coverage of
K-EUSO will be very important in understanding the nature of the
north-south anisotropy, or the inconsistency of the flux, being seen
by TA and Auger.

\subsection{A Future Ground Observatory}
Given that observations by the extended ground detectors will proceed
well into the next decade, and that exploratory space projects will
start giving a large acceptance coverage of the entire sky, what are
the new and/or remaining challenges for future ground observatories
(FGO) of ultra-high energy cosmic rays?  In this section, we take
``Auger$\times$10'' as a hypothetical example of an FGO, and discuss
how the FGO might look, and how research might proceed with the FGO.

{\bf FGO:} We assume ``Auger$\times$10'' is a symmetric set of
northern and southern observatories, each with 30,000 km$^2$ ground
area, covered in whole by hybrid arrays of FDs and SDs.  An array of
radio detectors (RDs) may be overlaid on the SD to enhance the energy
and possibly the \xmax{} determination for inclined events, improving
the quality of all sky coverage.  We assume the SD is equipped with a
particle identification  function for a fraction of shower
particles, and that this is to be used for the likelihood tagging of
the primary composition.  Approximately 10 \% of events are SD-FD
hybrid, which offers a direct means of composition determination via
\xmax{}.

{\bf Composition at the cutoff:} Such an FGO will collect
approximately 10,000 SD events above $E_{1/2}$ (10$^{19.8}$\,eV for
TA) or above $E_{s}$ (10$^{19.6}$\,eV for Auger) in 10 years of
operation, of which about 1,000 events will be SD-FD hybrid.  Protons
and iron are the natural nuclear species to compose a cutoff
structure, due to their expected abundance at the acceleration site
and their comparative stability in the subsequent propagation in the
nearby ($\sim$100\,Mpc) universe.  Indeed propagation calculations
indicate that cosmic rays above energies of $10^{19.5}$\,eV will have
a simplified, approximately bimodal arrival composition, even if they
are produced in equal proportions from protons to iron at the
source. Intermediate mass nuclei will appear mostly as proton and He
spallation by-products.  Thanks to the high statistics, the improved
\xmax{} resolution and the additional N$_\mu$ information of the FGO
hybrid events, we expect that contributions of protons and iron will
clearly stand out in a \xmax{} $-$ $N_\mu$ scatter plot.  Protons or
iron at the cutoff will be a straightforward confirmation of the
existence of the corresponding astrophysical mechanism that creates
the strong suppression, either the GZK or the acceleration limit
scenarios.

If protons and iron were both identified in the hybrid sample it would
allow the measured estimators of composition, \xmax{} and N$_\mu$,
and their predictions by the simulation code, to be ``calibrated'' by
the observation.  Even when contributions of He, CNO and heavier
nuclei are significant (and the isolation of protons and iron is not
obvious), we still expect proton and iron contribution, because the
existence of He results in (spallation) protons, and the existence of
CNO calls for the parent Fe of the spallation (see
Sections~\ref{Sect3:using_xmax} and \ref{Sect3:lack_of_iron} for a
discussion).  The statistics of the FGO hybrid sample, 1,000 events
above $E_{1/2}$ or $E_{s}$, would allow a reasonable ``calibration''
or cross-check to be performed for compositions in the range of
protons (Z=1) to iron (Z=26).

{\bf Composition dependent anisotropy and energy spectrum:} The SD and
RD events of the FGO are tagged with a likely primary mass derived
from the \xmax{} and N$_\mu$ analysis, both of which are being
calibrated using hybrid events.  The statistics of these events,
10,000 or more in total above the flux suppression, is enough to allow
the flux, energy spectrum and composition of UHECR to be separately
determined in $\sim$100 different sections of the sky. Their
correlations, such as the ``proton/iron sky above a certain energy''
and the ``energy spectrum of proton/He/CNO/iron in particular sections
of the sky", can be plotted from a single unified event sample. This
will be very effective in establishing astrophysical models to
explain the observed features of UHECR. Searches for auto-correlation
and association with astrophysical sources, as well as the
multi-messenger analysis, will be effectively made using the tag of
primary composition.

As a result, we can continue investigating the nature of galactic and
extra-galactic magnetic fields, background photons in the universe,
cosmological development of UHECR sources, special relativity with
exceptionally high Lorentz factors, and other subjects in astro-particle
physics.

{\bf Cosmogenic $\nu$s and $\gamma$s:} The search for UHE neutrinos
and gamma rays by the FGO will be limited only by statistics, using
the primary composition tagging of the FGO/SD.  The sensitivity to
cosmological neutrinos and gamma rays will allow us to enter the
region of possible detection, or of placing significant limits on
standard predictions (see Section~\ref{Sect2:photonsNeutrinos}).

{\bf UHE interactions:} Our understanding of ultra-high energy air
showers is incomplete; the data from present detectors indicate that
the number of shower particles in the off-core region is larger than
what the simulation program predicts, or that the simulated air shower
is ``slimmer'' than the real one (see
Section~\ref{Sect2:deficiencies}).  Using a large collecting area and
high sensitivity for penetrating particles, this is most clearly
demonstrated for muons detected in the Auger water-Cherenkov stations.  The
difference between the data and the simulation remains after the
ambiguities from energy determination and primary composition are
removed, and updated hadronic interaction models with the LHC data are
used~\cite{Auger_Hadron2016}.  Using composition-tagged SD events of
the FGO, the lateral distribution of muons and electrons in the
off-core region, and its relation to the primary energy and
composition will be studied.  The measurements are to be compared with
a variety of model predictions with the highest energy LHC data,
including taking into account nucleus-nucleus collisions.

In the case where a certain region of the sky is identified as
protonic without significant contributions of heavier components, the
ankle and the cutoff features in this region could possibly be
attributed to the pair-production on the CMB and the GZK effect, and
the corresponding energies can be used for calibrating the energy
scale of the incoming protons.  While it is possible that there may
still be a contribution from galactic protons, this can be checked by
examining the ratio of the ankle to GZK energies and the Berezinsky
modification factor~\cite{Aloisio2007} as a function of light/heavy
anisotropy. In any case, the possible anisotropies of these ratios
would be of great importance in constraining cosmic ray origin and
propagation models.  Given that these features are able to be
associated with pair production and the GZK cut-off, then the
difference between the expected primary energy and the measured
calorimetric energy by the FD or RD is to be accounted for by the
``invisible energy'' carried underground by the very high energy muons
and neutrinos in the shower core region.  In this way, we expect
ultra-high energy air showers will remain as a source of observational
information for the study of the nature of hadronic and nuclear
interactions beyond collider energies.

{\bf FD:} Measuring the energy and composition of UHECR will remain a
basic mission of the FGO.  The FGO/FD does this by covering the entire
acceptance, but with a limited duty cycle of $\sim$10 \%. The FGO/RD
would cover a limited acceptance at large $\theta$ but with a duty
cycle higher than 90\%, and its eventual contribution may become
significant.  When working in hybrid mode with the SD, some of the FD
information is redundant. This leads to the FAST concept of deploying
an array of compact, wide-angle and essentially single pixel FD
telescopes that record the time development of air fluorescence in
multiple stations~\cite{fast_design}.  The reconstruction of the
shower core location and arrival direction may be achieved mainly by
using the SD information, and the time variation of the FAST signal is
then converted into the longitudinal development of the shower.  The
FAST detector would work exclusively for supplying the calorimetric energy
and \xmax{} information of the event. Optimisation studies have been
performed with the goal of obtaining good resolution with limited
photon statistics.  Controlling the effect of background photons
on-line for the trigger and data acquisition remains a technical
challenge.  A small FD telescope with a similar concept has been
tested by the CRAFFT team~\cite{crafft_test}.  The current design
of FAST assumes FD stations with 360$^\circ$ azimuthal coverage on a
rectangular grid of 20\,km separation. A total of 75$\times$2 FD
stations will be necessary to cover the entire FGO acceptance in the
northern and southern hemispheres.

{\bf SD:} The FGO/SD is expected to have a good particle
identification capability for shower particles. The isolation of muons
will be of particular importance, and various types of detectors have
been tested during the design study of AugerPrime using the existing
water-Cherenkov station as a bulk muon counter and absorber of the EM
component. Here we remind the reader of another example, the ``lead
burger'': a sandwich of segmented scintillator and lead absorber,
tested in the AGASA array as a candidate for the original Auger SD
detector~\cite{hashimoto_icrc1995, lead_burger}.  Advances in
photo-detectors and electronics may now allow a significantly finer
detector segmentation and fast waveform sampling, strengthening the
multi-hit capability of the lead burger~\cite{mu_ray,
  nonaka_uhecr2013, peters_uhecr2016}.  dE/dX measurement and coarse
tracking of individual particles may also be incorporated. Besides
muon identification, detection of spallation neutrons may be possible
for tagging the nuclear composition and identifying primary gamma
rays.  TA and Auger are ideal testing grounds for developing the
FGO/SD and for optimising its performance.  Taking the 2.1km grid
spacing of TA$\times$4 as an example, 6,900$\times$2 SD units will be
necessary to fill the FGO acceptance.

{\bf Electronics and Network:} The FGO electronics may follow the base
design of Auger and TA with FADC sampling of wave-forms, multi-level
digital triggers and wireless communication networks. For the FGO/SD,
the number of readout channels must be significantly increased for the
segmented detector and for the integration of FGO/RD, and faster
digitisation is required for better timing measurement. The biggest
challenge would be that all these performance upgrades need to be
realised with a limited power budget due to local electricity
generation and storage.  Taking advantage of the low duty cycle of SD
digitisation electronics, clever methods could be invented to save
electric power consumption.

Reliability and fault-tolerance are required for the stable operation
of many FGO/SDs distributed over a large ground area.  With a steady
increase of locally available computing and data storage capacities,
real-time requirements for trigger generation and data acquisition may
be loosened, and greater autonomy may be allowed for the operation of
individual SDs. This in turn reduces the network load of
communications between SDs, and increases the overall system
reliability.  It may also allow for the whole communication system of
the FGO to reside on a standard communication network. This will be
advantageous in terms of construction cost, long-term operation and
for taking advantage of progress in communication technologies and
updates in the DAQ system.  Implementing a prompt coincidence trigger
formed over several clustered SDs, assuming it is required, may be a
technical challenge in building such an autonomous DAQ system.

For the FGO/FD array, the load on digitisation electronics will
decrease compared with the current many-channel system, but triggering
on a limited number of sensor pixels will be itself a challenging
task. A clever time coincidence method between the neighboring SDs and
FDs, or the introduction of an external trigger via the network, may
be a resolution to this problem. Reliable remote operation of the
telescopes, monitoring and calibration devices via the wireless
network is the key to the success of achieving good operational
performance of the FGO.

{\bf Collaboration:} The construction and operation of the FGO will be
a challenge for technology, management and resources.  It may be
accomplished only through a collaboration of people with zeal, having
excellent expertise and experience.  Some of the features of the FGO
detector, electronics and communication system were already discussed
in the proposals for northern Auger and AugerPrime~\cite{AugerNorth,
  AugerPrime}.  The physics issues, and an experiment of similar
sensitivity to the FGO known as ``TA2'', have been discussed in
physics community meetings in Japan.

\section{Conclusion}
The era of the very large observatories has produced many new and
important insights into the properties of the ultra-high energy cosmic
ray flux.  This has been achieved with well designed detectors and
very large collecting areas, with important design input from previous
generations of experiments.  An important feature of both Auger and TA
is the hybrid nature of the observations.  Apart from providing
calorimetric energy measurements, a hybrid observatory offers a
multitude of cross-checks which have improved the measurements and
reduced the systematic uncertainties.

The UHECR energy spectrum is now measured with high statistics,
resulting from a combined Auger/TA exposure of over
60,000\,km$^2$\,sr\,yr at the highest energies.  The spectrum now
reveals several features including an unambiguous suppression beyond
$4\times10^{19}$\,eV.  A dipole anisotropy has been observed for the
first time at ultra-high energies, and there is great interest in the
possible northern hemisphere ``hot spot'' over a $20^\circ$ radius
area of sky which will be monitored by TA$\times$4 for an increase in
significance in coming years.

Small-scale anisotropies and associations of cosmic ray arrival
directions with astronomical catalogs have not been convincingly
observed.  It is probably fair to say that this lack of success was
not expected.  This may be related to an apparent increase in the
cosmic ray mass and charge above the ankle of the energy spectrum.
While not currently embraced by the entire community, a heavier flux
would help to explain the lack of small-scale anisotropy.  On the
other hand, if the northern hemisphere ``hot spot'' persists, its
appearance may be related to mass composition differences between the
north and the south.  The difference of the flux suppression energies
between Auger and TA may also suggest that the astrophysics is not identical
in both hemispheres.  In that context, a composition dependent
anisotropy study will be of great interest.  The promising aspect is
that our measurements of air shower development, whether they be from
FDs or SDs, are continuing to improve, with the likely future outcome
being the ability to (at least) identify the lighter fraction of the
flux with a surface detector, 24 hours per day, as being realised by
AugerPrime.

Photon and neutrino limits set by the experiments have ruled out
certain exotic production mechanisms for the highest energy cosmic
rays, including the decay of super-heavy dark matter.  They are now
probing cosmogenic photon and neutrino models that may provide
information on the fraction of protons present at the highest
energies.  Finally, measurements at the observatories are constraining
some aspects of hadronic and nuclear interaction models at very high
energy, a remarkable feat given the indirect view of the interactions
afforded by characteristics of the air shower.

As well as these achievements, we have also discussed the present
challenges for the field.  For example, despite much progress in
measurements at accelerators, and despite constraints from
ultra-energetic cosmic rays, we do not know the systematic uncertainty
to attach to a simulation prediction for the mean \xmax{} of a proton
(or any) shower, reducing the power of our mass measurements and their
influence on the open astrophysics questions.  It also means that we
are currently limited in being able to confidently select showers
initiated by low charge primaries in our attempts to improve the
sensitivity to anisotropies, especially since we do not yet have a
mass estimate for every event we observe.  Finally, we still lack
sufficient collecting area to answer some of the big questions, given
that we appear to be faced with a cosmic ray sky remarkably free of
strong anisotropies at the highest energies.

For the future, we must build on this impressive progress with new
ideas and techniques that will lead to new observatories that are
sufficiently large (and cost-effective) to answer the remaining
questions.  While further increases in collecting area are of prime
importance, it seems crucial that future ground observatories be endowed with
at least some mass-composition sensitivity for all collected events.

\section*{Acknowledgment}

We thank all of our colleagues in the Pierre Auger and Telescope Array
experiments for their inspiring collaboration.  While we have
benefitted greatly from discussions with many of them over the years,
the responsibility for any errors in this article is entirely ours.


\bibliographystyle{ptephy_BRD}


%

\end{document}